\documentclass[letterpaper,twocolumn,10pt]{article}
\usepackage{makecell}  
\usepackage{booktabs} 
\usepackage{amsmath}
\usepackage{enumitem}
\usepackage{hyperref}
\usepackage{url}
\usepackage{tabularx}
\usepackage{booktabs}
\usepackage{cleveref}
\usepackage[frozencache,cachedir=.]{minted}
\usepackage[normalem]{ulem}

\usepackage{xcolor}
\usepackage[table]{xcolor} 
\newcommand{\best}[1]{\cellcolor{blue!10}{\textbf{#1}}}
\newcommand{\bestnobold}[1]{\cellcolor{blue!10}{#1}}
\newcommand{\nbc}[3]{
 {\colorbox{#3}{\bfseries\sffamily\scriptsize\textcolor{white}{#1}}}
 {\textcolor{#3}{\sf\small\textit{#2}}}
 }

\definecolor{kscolor}{rgb}{0.9,0.1,0.1}
\definecolor{mscolor}{rgb}{0.1,0.1,0.9}
\definecolor{stcolor}{rgb}{0.1,0.9,0.1}

\usepackage{subcaption}
\usepackage{titling}

\definecolor{shubhambrown}{RGB}{150, 75, 0} 

\newif\ifcomments
\newcommand{\eat}[1]{}
\commentstrue
\ifcomments
    \providecommand{\shu}[1]{{\color{orange}{/* shu: #1 */}}}
    \providecommand{\melissa}[1]{{\color{magenta}{/* melissa: #1 */}}}
    \providecommand{\accheng}[1]{{\color{blue}{/* accheng: #1 */}}}
    \providecommand{\shubham}[1]{\textcolor{shubhambrown}{{[Mert/Shubham: #1]}}}
    \providecommand{\ion}[1]{{\color{cyan}{/* ion: #1 */}}}
    \providecommand{\jx}[1]{{\color{red}{/* xing: #1 */}}}
    \providecommand{\andyl}[1]{{\color{brown}{/* andyl: #1 */}}}
    \providecommand{\tian}[1]{{\color{teal}{/* tian: #1 */}}}
    \providecommand{\bowen}[1]{{\color{olive}{/* bowen: #1 */}}}
    \providecommand{\mert}[1]{{\color{olive}{/* mert: #1 */}}}
    \providecommand{\red}[1]{{\color{red}{/*#1 */}}}
    \providecommand{\ks}[1]{\nbc{KS}{#1}{kscolor}}
    \providecommand{\todo}[1]{
    {\colorbox{red}{\bfseries\sffamily\scriptsize\textcolor{white}{TODO}}}
     {\textcolor{red}{\sf\small\textit{#1}}}
    }
    
\else
    \providecommand{\shu}[1]{}
    \providecommand{\melissa}[1]{}    
    \providecommand{\accheng}[1]{}
    \providecommand{\shubham}[1]{}
    \providecommand{\ion}[1]{}
    \providecommand{\jx}[1]{}
    \providecommand{\andyl}[1]{}
    \providecommand{\tian}[1]{}
    \providecommand{\bowen}[1]{}
    \providecommand{\mert}[1]{}
    \providecommand{\red}[1]{}
    \providecommand{\ks}[1]{}
    \providecommand{\todo}[1]{}
    
\fi

\definecolor{takeawaybackground}{HTML}{F7FAFC} 
\definecolor{takeawayborder}{HTML}{6B7280}     

\usepackage[most]{tcolorbox}

\newcommand{\takeaway}[2]{%
  \par\vspace{0.35\baselineskip}%
  \noindent
  \begin{tcolorbox}[
    colback=takeawaybackground,
    colframe=takeawayborder,
    boxrule=0.9pt,
    arc=1mm,            
    left=6pt,
    right=6pt,
    top=6pt,
    bottom=6pt,
    boxsep=0pt,
    enhanced,
  ]
  \noindent\textbf{Takeaway #1.}~#2
  \end{tcolorbox}
  \par\vspace{0.35\baselineskip}%
}

\newcommand{\SYS}{}
\def\SYS/{ADRS}
\newcommand{\NumCase}{}
\def\NumCase/{ten}
\newcommand{\numcase}{}
\def\numcase/{10}
\usepackage{tabularx} 
\usepackage{makecell} 
\usepackage{array}    
\usepackage{graphicx} 
\usepackage{tikz}
\usepackage{multirow}
\usepackage{tabularx} 
\usepackage{xcolor}
\usepackage{pifont}
\usepackage{tcolorbox}
\usepackage{enumitem}
\setitemize{noitemsep,topsep=0pt,parsep=0pt,partopsep=0pt}

\usepackage{listings}
\usepackage{xcolor}

\usepackage{preamble}

\newtcolorbox{takeaway2box}[1][]{%
  enhanced,
  sharp corners,
  colback=takeawaybackground,
  colframe=takeawayborder,
  boxrule=0.5mm,
  left=1mm, right=1mm, top=1mm, bottom=1mm,
  before skip=5pt, after skip=5pt,
  #1
}

\begin{document}
\date{}

\title{Let the Barbarians In:\\ How AI Can Accelerate Systems Performance Research}

\author{
\begin{tabular}{c}
{\normalfont Audrey Cheng\thanks{Equal contribution, ordered alphabetically.}, \, Shu Liu\textsuperscript{*}, Melissa Pan, Zhifei Li, Shubham Agarwal, Mert Cemri,}\\
{\normalfont Bowen Wang, Alexander Krentsel, Tian Xia, Jongseok Park, Shuo Yang, Jeff Chen, }\\
{\normalfont Lakshya Agrawal, Ashwin Naren, Shulu Li, Ruiying Ma, Aditya Desai, Jiarong Xing, }\\
{\normalfont Koushik Sen, Matei Zaharia, Ion Stoica}
\end{tabular}\\[1.5em]
UC Berkeley}


\maketitle

\subsection*{Abstract}
Artificial Intelligence (AI) is beginning to transform the research process by automating the discovery of new solutions. This shift depends on the availability of reliable verifiers, which AI-driven approaches require to validate candidate solutions. Research focused on improving systems performance is especially well-suited to this paradigm because system performance problems naturally admit such verifiers: candidates can be implemented in real systems or simulators and evaluated against predefined workloads. We term this iterative cycle of generation, evaluation, and refinement AI-Driven Research for Systems (ADRS). Using several open-source ADRS instances (i.e., OpenEvolve, GEPA, and ShinkaEvolve), we demonstrate across ten case studies (e.g., multi-region cloud scheduling, mixture-of-experts load balancing, LLM-based SQL, transaction scheduling) that ADRS-generated solutions can match or even outperform human state-of-the-art designs. Based on these findings, we outline best practices (e.g., level of prompt specification, amount of feedback, robust evaluation) for effectively using ADRS, and we discuss future research directions and their implications. Although we do not yet have a universal recipe for applying ADRS across all of systems research, we hope our preliminary findings, together with the challenges we identify, offer meaningful guidance for future work as researcher effort shifts increasingly toward problem formulation and strategic oversight. 

\emph{Note: This paper is an extension of our prior work~\cite{cheng2025barbarians}. It adds extensive evaluation across multiple ADRS frameworks and provides deeper analysis and insights into best practices.}

\section{Introduction}

One of the most ambitious goals of artificial intelligence (AI) is to revolutionize scientific discovery by automating algorithm design, experiment execution, and even the research process itself. While the realization of this goal will likely be uneven, with certain domains being transformed earlier and more profoundly than others, AI-driven approaches~\cite{alphaevolve,openevolve,agrawal2025gepa,shinkaevolve,hamadanian2025glia} have already reached a level of capability where they can meaningfully contribute to computer systems research. 

A significant portion of systems research, spanning networking, databases, and distributed systems, is dedicated to enhancing performance. This is typically achieved through the meticulous, human-driven design of new algorithms for tasks such as routing~\cite{Boukerche-2011-ComputerNetworks, SirikaMahajan2016SurveyDynamicRouting, Suryanar23-Routing}, scheduling~\cite{wu2024can,cheng2024towards}, and resource management~\cite{Gao_2024,KhanPasrichaKim2020_PIMNMP_Survey}. Crucially, the novelty and efficacy of these algorithms are often the primary metrics for a publishable paper. 

Our first inquiry in this paper is to explore whether a new class of AI-driven approaches, which we term AI-Driven Research for Systems (\SYS/), can generate algorithms that surpass human state-of-the-art solutions for systems performance problems. To answer this inquiry, we use three emerging open-source \SYS/ frameworks (e.g., OpenEvolve~\cite{openevolve}, GEPA~\cite{agrawal2025gepa}, and ShinkaEvolve~\cite{shinkaevolve}) across ten real research tasks and show that these frameworks can already generate solutions that match or even exceed the performance of state-of-the-art, human-designed solutions. For example, in a load-balancing problem for a Mixture-of-Experts (MoE) model, OpenEvolve discovers an algorithm to rebalance experts across GPUs that is 13$\times$ faster than the best-known baseline. In a job scheduling problem aimed at reducing costs by using spot instances across multiple cloud regions, OpenEvolve generates a solution that achieved roughly 35\% greater savings than an expert-developed baseline (Table \ref{tab:project-summary}). 

Given these compelling results, we turn to the question of how to best apply these frameworks for solution discovery. We conduct extensive ablation studies across all components of the evolutionary process and based on these findings, outline best practices along three axes: \textit{problem specification}, \textit{solution evaluation}, and \textit{feedback}. For specification, we find that ``less is more and more is less'': the amount of information we provide is critical to navigate the trade-off between exploration and exploitation, i.e., between quickly finding a solution and spending additional time and resources searching for a better solution. For evaluation, we find that the ``generated solutions are only as strong as their evaluators'': discovering effective solutions requires ensuring robust verification and diverse test sets. Finally, we show that for feedback, the ``devil is in the details'': maximizing performance depends on calibrating feedback granularity to provide actionable guidance without overfitting. 



Finally, we discuss the consequences of AI-driven research for the systems community. As AI increasingly takes on the role of algorithm discovery and optimization, the emphasis for human researchers will likely pivot to problem formulation, high-level ideation, and strategic direction. In this new model, the researcher acts as an advisor to powerful AI research assistants: defining meaningful problems, proposing creative starting points, and distilling insights from generated solutions. This approach can create a powerful virtuous cycle: the same AI-driven methodologies can be applied to improve the AI systems themselves, leading to a compounding acceleration of the pace of discovery.

\eat{
\begin{figure}[h!]
  \centering
  \vspace{-0.7em}
  \begin{subfigure}[t]{0.48\textwidth}
    \centering
    \includegraphics[width=\textwidth]{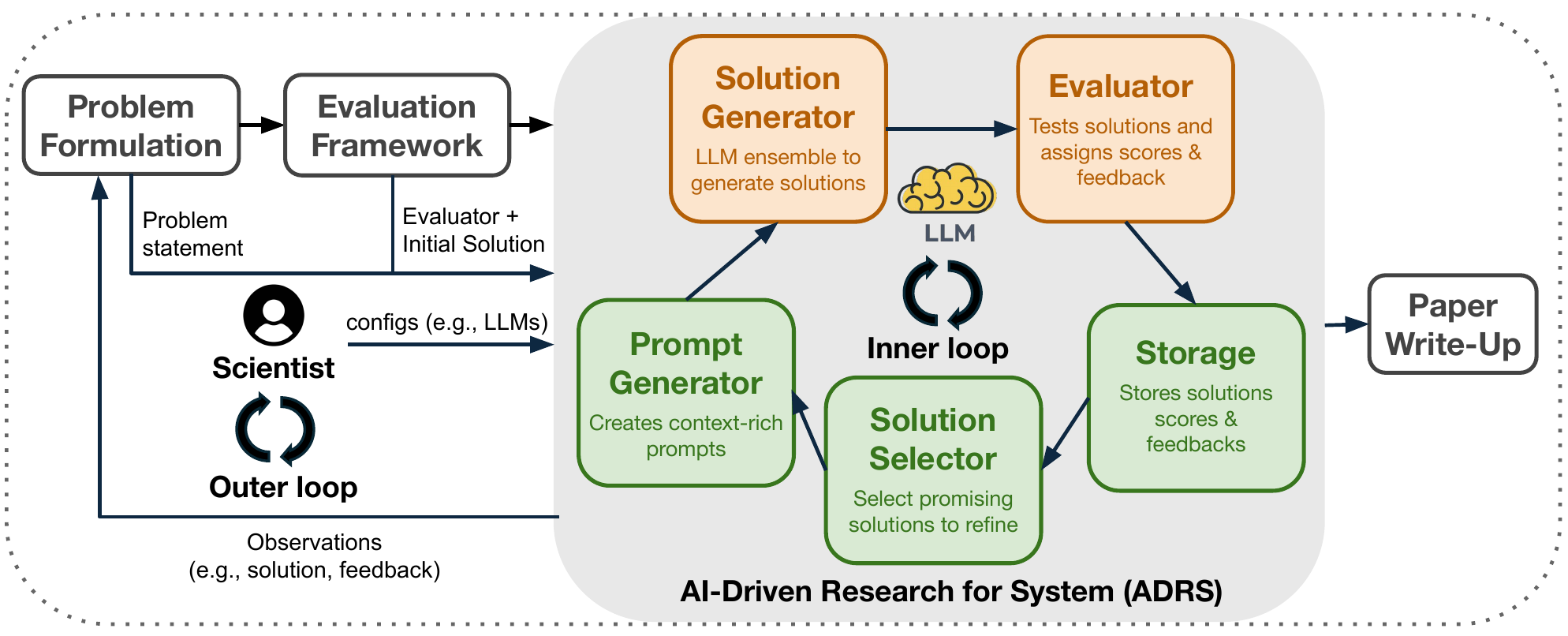}
    \caption{AI-Driven Research for Systems architecture.} 
    \label{fig:evolve-sys}
  \end{subfigure}%
  \hfill
  \begin{subfigure}[t]{0.48\textwidth}
    \centering
    \includegraphics[width=\textwidth]{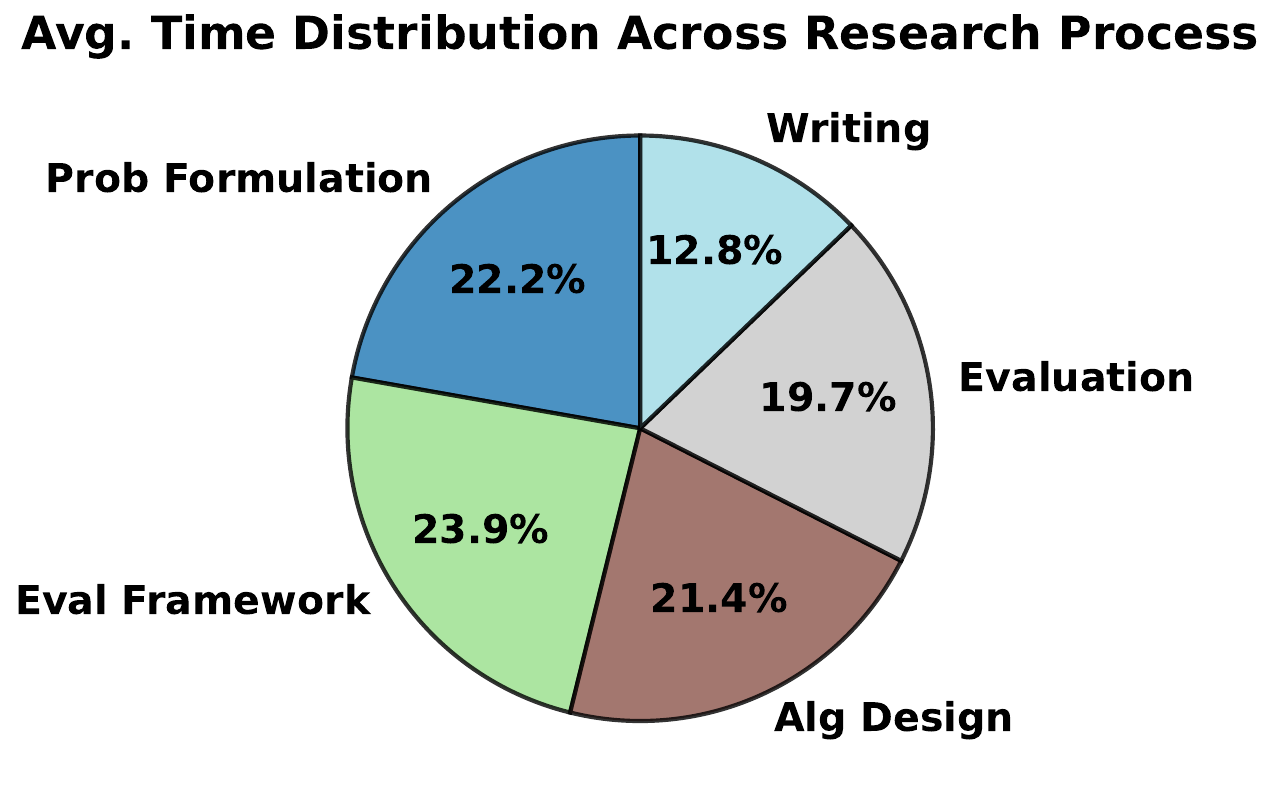}
    \caption{Research Process Survey.} 
    \label{fig:alphaevolve}
  \end{subfigure}
  \vspace{-0.5em}
  \caption{Comparison of system architecture (left) and survey process (right).}
\end{figure}
}

\begin{figure*}[t!]
\centering
\begin{subfigure}[t]{0.5\textwidth}
    \vspace{0pt} 
    \centering
    \includegraphics[width=\linewidth]{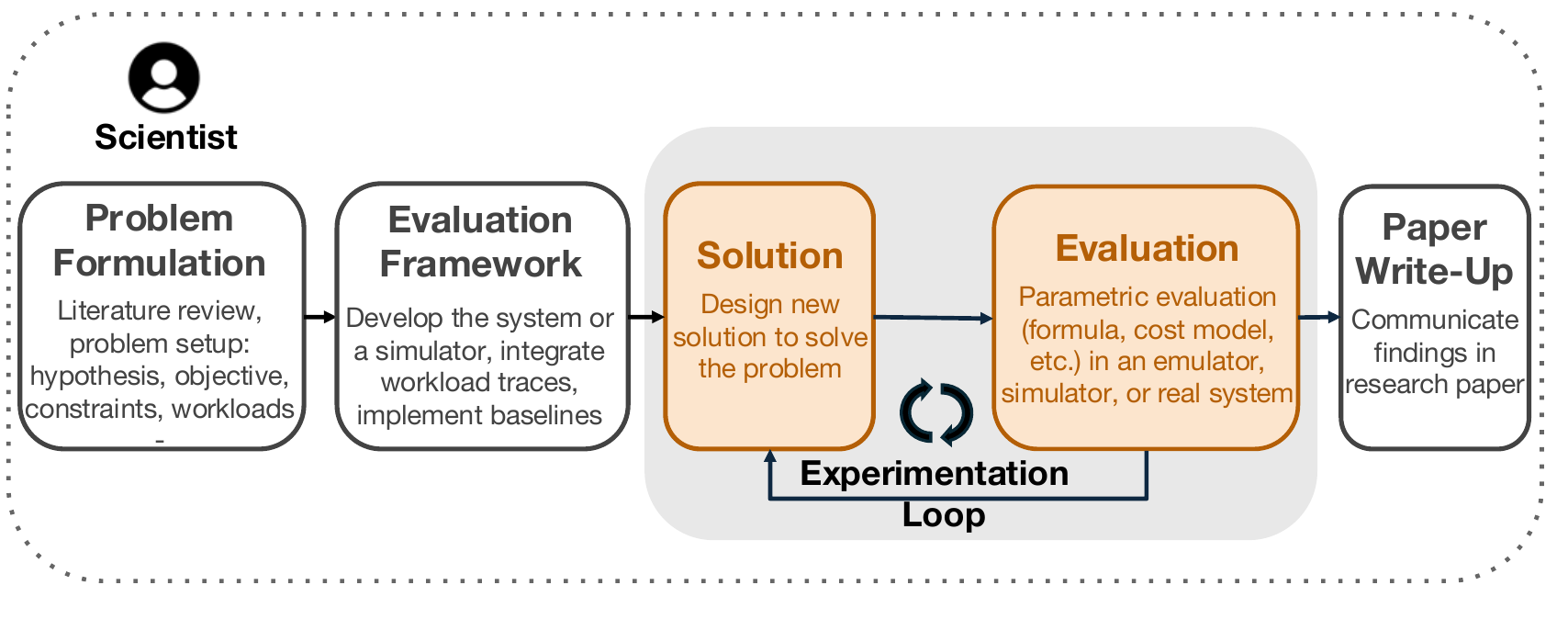}
    \caption{\phantom{(a)}} 
    \label{fig:research-process}
\end{subfigure}%
\hfill
\begin{subfigure}[t]{0.5\textwidth}
    \vspace{0pt} 
    \centering
    \includegraphics[width=\linewidth]{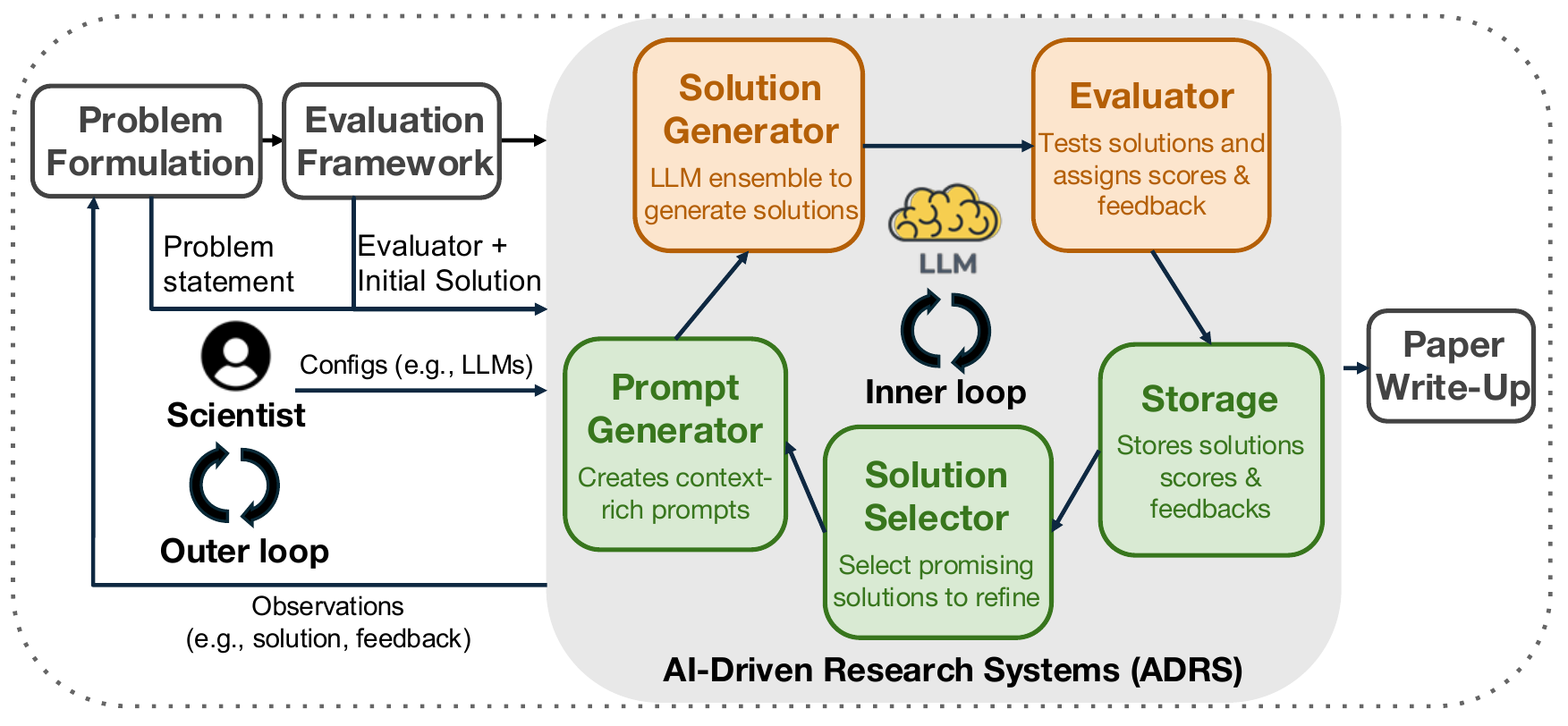}
    \caption{\phantom{(b)}} 
    \label{fig:evolve-sys}
\end{subfigure}
\vspace{-0.5em}
\caption{
(a) The five stages of a typical systems research process. 
(b) The AI-driven Research for Systems (ADRS) architecture instantiated in the same process. 
ADRS (Grey area) augments the scientist by automating the \textbf{Solution} and \textbf{Evaluation} stages while leaving the other stages unchanged.}
\label{fig:research-process-and-evolve-sys}
\end{figure*}

In summary, we introduce no new problems, algorithms, or mechanisms in this paper. Instead, we evaluate how emerging AI-driven approaches may substantially accelerate systems research and articulate why embracing this approach may be essential as the field evolves. We make three contributions:
\begin{itemize}
    \item We study the capabilities of existing open-source \SYS/ frameworks for systems performance problems across ten case studies, showing that AI-generated algorithms can outperform human state-of-the-art solutions (e.g., 13$\times$ faster load balancing and 35\% lower cloud costs).
    \item We suggest a set of best practices for problem specification, evaluation, and feedback to help researchers effectively apply \SYS/.
    \item We discuss the implications of this shift for the systems community, outlining the evolving role of researchers, and identifying open challenges in AI-assisted discovery.
\end{itemize}

\section{Why AI-Driven Research for Systems?}
\label{why-systems}


In this paper, we advocate for an AI-driven approach to systems performance problems.
While performance optimization is not the sole focus of systems research, it remains a central one—a brief survey of top systems, networking, and database venues (NSDI, OSDI, SIGMOD, SOSP, and VLDB) shows that over one-third of published papers feature performance optimization algorithms as their core contribution. We argue that AI-driven methods are particularly well-suited to this domain because candidate solutions can be verified robustly and at low cost.

First, verifying whether a solution improves performance is straightforward. Such solutions typically introduce new techniques or algorithms implemented directly within the systems they aim to optimize. Verification amounts to running these systems under representative workloads and checking for improvements over baselines on relevant performance metrics.

Second, solutions to systems performance problems typically preserve the correctness of the original system, or at least, verifying this is straightforward. For instance, it is straightforward to check whether a load-balancing algorithm schedules all assigned tasks or whether a network router forwards all packets it receives.

Third, the algorithmic code targeted for evolution is often relatively small, e.g., the core logic of a scheduler, load balancer, or resource allocator. This makes the generated code easier for humans to interpret, which can further help with correctness verification. In all our case studies, we readily understood the generated solutions and identified their main ideas and techniques (see Section~\ref{sec:case_studies}). As these tools mature, we expect the scope of modification to expand across multiple components (e.g., when designing complex distributed protocols). Maintaining code interpretability in such cases is an important topic for future research.

Finally, systems researchers often use simulators to develop and evaluate solutions before deployment. Since simulator-based verification is relatively cheap, evaluation remains practical even if the search process generates excessive candidate solutions. For example, most of our case studies required only a few hours and cost less than several tens of dollars. That said, building inexpensive yet faithful simulators for complex systems (e.g., operating systems, databases) is far from trivial and remains a topic for future research (Section~\ref{sec:better-evaluators}).

\section{Using AI to Accelerate Systems Research}

This section provides an overview of the systems research process and then introduces the AI-Driven Research for Systems (\SYS/) approach, which accelerates this process through automatic solution discovery and evaluation.

\subsection{Systems Performance Research Process}

The typical systems research process spans weeks or months and consists of five stages (Figure~\ref{fig:research-process}):

\begin{itemize}

\item \textbf{Problem Formulation:} Precisely define the problem to solve (e.g., improving system throughput). This problem anchors the entire research process and the final results.

\item \textbf{Evaluation Framework:} Develop a framework to implement and evaluate potential solutions. This may be the system itself or a simulator that approximates the system's behavior. Even when the system exists, a simulator may be used to accelerate iteration. Researchers also assemble or reuse workloads (traces) or benchmarks to drive evaluation.

\item \textbf{Solution:} Design a solution (e.g., an algorithm) for the problem, such as a new scheduling or search technique.

\item \textbf{Evaluation:} Implement the solution in the simulator or system, evaluate its performance using the selected workloads, and compare it against baseline(s). If the solution does not improve performance, researchers return to the \textbf{Solution} stage to refine the approach or develop alternatives.

\item \textbf{Paper Write-Up:} Once a solution achieves the desired results, document and communicate the findings.

\end{itemize}

\eat{
\begin{figure*}[t]
\centering
\begin{subfigure}[t]{0.5\textwidth}
    \vspace{0pt} 
    \centering
    \includegraphics[width=\linewidth]{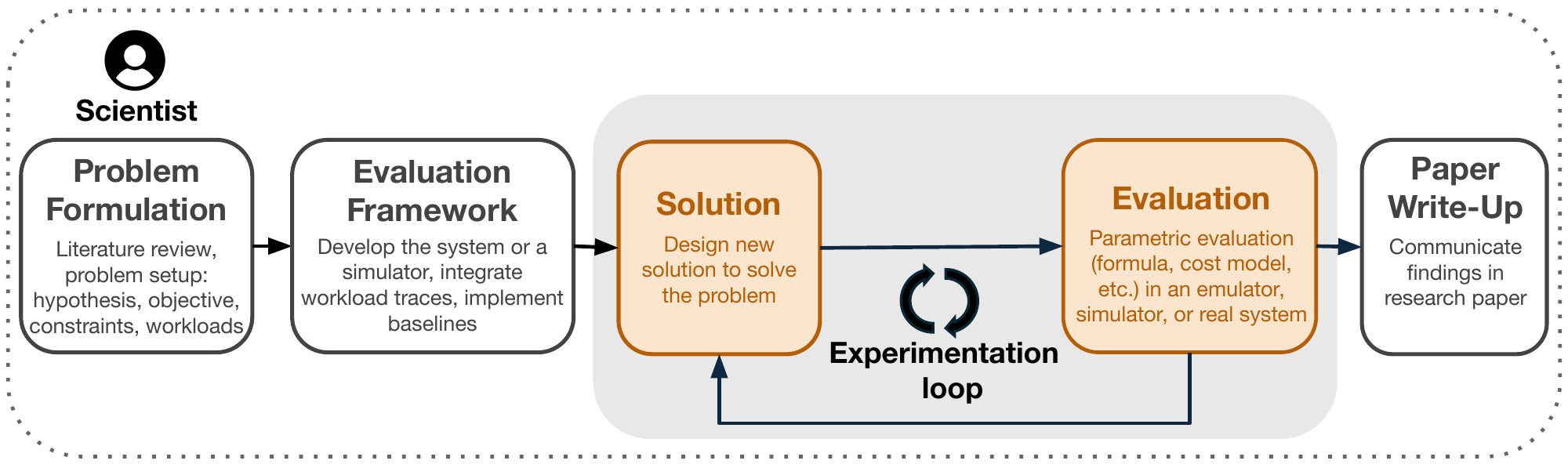}
    \caption{\phantom{(a)}} 
    \label{fig:research-process}
\end{subfigure}%
\hfill
\begin{subfigure}[t]{0.5\textwidth}
    \vspace{0pt} 
    \centering
    \includegraphics[width=\linewidth]{figures/evolve-sys.pdf}
    \caption{\phantom{(b)}} 
    \label{fig:evolve-sys}
\end{subfigure}

\vspace{-0.5em}
\caption{
(a) The five stages of a typical systems research process. 
(b) The AI-driven Research for Systems (ADRS) architecture instantiated in the same process. 
ADRS (Grey area) augments the scientist by automating the \textbf{Solution} and \textbf{Evaluation} stages while leaving the other stages unchanged.
}
\label{fig:research-process-and-evolve-sys}
\end{figure*}
}

\subsection{AI-Driven Research for Systems (\SYS/)}
\label{subsec:sys}
As Large Language Models (LLMs) have progressed from simple text completion to sophisticated reasoning and tool use, new architectures have been developed to enhance the reliability and scope of tasks to which they can be applied. In this paper, we focus on how LLMs can be used to design, implement, and evaluate new solutions (e.g., algorithms) to solve systems research problems. We call this approach AI-Driven Research for Systems (\SYS/) and depict it in Figure~\ref{fig:evolve-sys}. \SYS/ operationalizes the two iterative stages of the systems research process—\textbf{Solution} and \textbf{Evaluation}—shown in Figure~\ref{fig:research-process}. Together, these stages account for roughly 40\% of total research time (based on a survey of over 30 systems researchers, Figure~\ref{fig:research-process-survey} in 
Appendix~\ref{app:survey_result}).

At its core, \SYS/ implements an iterative loop that constructs or refines prompts for LLMs, generates new solutions or improves existing ones, and evaluates these solutions in either a real system or a simulator. This loop continues until a satisfactory solution emerges, the resource budget is exhausted, or the researcher terminates it. \SYS/ consists of five components:

\begin{itemize}

\item \emph{Prompt Generator}: Creates the prompt used to generate the solution. This prompt contains the problem statement, context (e.g., simulator or system code), and previous solutions (from the \emph{Solution Selector}) to guide refinement.

\item \emph{Solution Generator}: Feeds the prompt from the \emph{Prompt Generator} to one or more LLMs to generate a new solution or refine an existing one, typically by directly editing the code in the simulator or real system.

\item \emph{Evaluator}: Takes the solution from the \emph{Solution Generator} and runs it against a predefined set of workloads. The solution is scored based on performance and can use an LLM to provide qualitative feedback. 

\item \emph{Storage}: Persists solutions, outputs, scores, and the feedback provided by the \emph{Evaluator}.

\item \emph{Solution Selector}: Chooses a subset of solutions from \emph{Storage} and provides them to the \emph{Prompt Generator} to seed the next generation.

\end{itemize}

Together, these components form an automated inner feedback loop that enables \SYS/ to iteratively refine solutions. This can be paired with an outer loop in which a human can observe the generated solutions and provide high-level guidance for future prompts.

In most cases, \SYS/ leverages simulators rather than real systems for two main reasons. First, the real system's codebase is often too large to fit within the context window of current LLMs. Second, running evaluations in a simulator can be orders of magnitude faster than on the real system, greatly accelerating iteration.




\begin{table*}[t]
\small
\setlength{\tabcolsep}{6pt}
\centering
\begin{tabularx}{\textwidth}{lXXX}
\toprule
\textbf{Framework} & \textbf{Parent Selection} & \textbf{Evolution Context for Generation} & \textbf{Evaluation \& Cost} \\
\midrule
OpenEvolve &
Sample a parent per island using exploration, archive-based exploitation, or random selection. &
Parent program plus a few top and diverse archive programs from the same island. &
Run full validation; higher-scoring program is added to the archive. One LLM call per iteration; no correctness gate. \\
\midrule
GEPA &
Sample parents from a per-instance Pareto frontier over validation instances. &
Parent code together with minibatch evaluation traces and metrics. &
First require improvement on a small minibatch, then run full validation. Approximately one LLM call and 6--11 evaluations per iteration. \\
\midrule
ShinkaEvolve &
Weighted sampling from a Pareto archive, with bonuses for novelty and programs with fewer children. &
Parent plus six archive inspirations, combined with meta-recommendations based on program summaries. &
Only correct programs are used for evolution. 
One LLM plus one embedding call per iteration. 12 LLM calls for meta-learning every 10 iterations. \\
\bottomrule
\end{tabularx}
\caption{\textbf{ADRS frameworks we evaluate in our experiments.}
All use the same island-based population structure; we highlight only the design choices and costs relevant to our evaluation.}
\label{tab:framework-overview}
\end{table*}

As mentioned earlier, \SYS/ automates the \textbf{Solution} and \textbf{Evaluation} stages of the research process depicted in Figure~\ref{fig:research-process}. 
As such, \SYS/ can accelerate two of the most time-consuming stages (represented in orange in Figure~\ref{fig:research-process}): \textbf{Solution Development} and \textbf{Evaluation} by iteratively proposing and refining solutions until a satisfactory one is discovered or the iteration limit is reached, optionally incorporating researcher guidance.

In addition to speed, we believe \SYS/ is valuable because it transcends the domain-specific expertise of human researchers. LLMs are trained on vast, diverse datasets and can discover novel solutions by identifying patterns from unrelated domains--solutions that a human expert might simply overlook due to their specialized focus. In our experiments, LLMs combined techniques from diverse domains such as Political Science and Economics to outperform human SOTA solutions (Table~\ref{tab:cross-domain-techniques}, Appendix~\ref{appendix:cross-domain}). For example, for the MoE load balancing algorithm (\textbf{EPLB}), GEPA with GPT-5 generated a solution that uses Hamilton's Apportionment from Political Science to efficiently assign replica counts for uniform workloads without expensive search iterations.

Finally, it is important to note that for other stages of the research process, \SYS/ has a neutral effect: it makes it neither harder nor easier for researchers to choose problems, build evaluation frameworks, or document results. Thus, \SYS/ advances the Pareto frontier of the research process.

\subsection{\SYS/ Examples}
\label{sec:adrs-examples}

\SYS/ aligns with several recent systems that pair LLM-generated candidates with programmatic evaluation. AlphaEvolve~\cite{alphaevolve} is a proprietary ADRS framework from Google DeepMind that discovers new algorithms using an evolutionary loop over LLM-generated candidates managed via MAP-Elites and an island model.

We evaluate three open-source ADRS frameworks: OpenEvolve~\cite{openevolve}, GEPA~\cite{agrawal2025gepa}, and ShinkaEvolve~\cite{shinkaevolve} (Table~\ref{tab:framework-overview}). OpenEvolve re-implements AlphaEvolve’s core asynchronous pipeline and adds developer utilities, such as evolution tree visualization. GEPA takes a prompt-centric approach, using natural-language reflection to mutate prompts and Pareto filtering to retain diverse high-performing solutions. ShinkaEvolve emphasizes structured introspection through weighted archive sampling, a correctness gate that admits only valid programs, and periodic meta-reflection. 

Coding assistants, such as  Claude Code~\cite{ClaudeCode2025} and Cursor~\cite{CursorAgent2024}, can also be viewed as \SYS/ instances that leverage codebase context and natural-language prompts to evolve and verify algorithms through an interactive feedback loop.

\section{Evaluation and Case Studies}
\label{sec:case_studies}

\begin{table*}[t]
\centering
\footnotesize
\setlength{\tabcolsep}{4pt}
\renewcommand{\arraystretch}{1.2}
\setlength{\extrarowheight}{1pt}

\begin{tabularx}{\textwidth}{@{}p{3.5cm} p{6cm} X p{2.5cm}@{}}
\toprule
\textbf{Task \& SOTA Publication} & \textbf{Objective} & \textbf{SOTA / Baseline} & \textbf{Time / Cost} \\
\midrule

Telemetry Repair \cite{krentsel2024case} &
Repair buggy network telemetry. &
+11.8\% better counter repair score, +47.9\% higher confidence calibration. &
8h (100 iters), $\leq\$15$ \\
\midrule


Cloudcast \cite{wooders2024cloudcast} &
Optimize multi-cloud data transfer cost. &
No improvement. & 
1h (100 iters), $\leq\$15$ \\
\midrule

Expert Parallelism Load \newline Balancer~\cite{deepseek-eplb} (EPLB) &
Balance expert-parallel load across GPUs. &
Same load balance, 13$\times$ faster runtime vs.\ proprietary implementation. &
5h (100 iters), $\leq\$15$ \\
\midrule

Model Placement \cite{yu2025prism} (Prism) &
Optimize cost for model-to-GPU placement globally. &
20.9\% cheaper than published solution. &
1h (100 iters), $\leq\$15$ \\
\midrule

LLM-SQL \cite{liu2024optimizing} &
Reorder tabular data to improve prefix hit rate. &
Comparable hit rate, 3.9$\times$ faster runtime. &
1h (100 iters), $\leq\$20$ \\
\midrule

Transaction Scheduling \cite{cheng2024towards} (TXN) &
Minimize makespan in transaction scheduling. &
60\% better than greedy (offline). &
$<$2h (100 iters), $\leq\$20$ \\
\midrule

Can’t Be Late \cite{wu2024can} (CBL) &
Schedule deadline-driven jobs on single-region spot instances. &
Up to 16\% (average 7\%) higher cost savings vs.\ SOTA. &
5h (100 iters), $\leq\$30$ \\
\midrule

Can’t Be Late Multi-Region Extension [WIP]  (CBL-Multi) &
Schedule deadline-driven jobs on multi-region spot instances. &
26\% lower cost vs.\ single-region baseline. &
3h (100 iters), $\leq\$25$ \\
\midrule


Multi-Agent System Optimization \cite{metagpt} (MAS) &
Improve multi-agent collaboration using MAST taxonomy. &
7\% improvement on ProgramDev. &
$<$2h (100 iters), $\leq\$15$ \\
\midrule

Datacenter TCP Congestion Control \cite{powertcp} (NS3)&
Maximize throughput and minimize queue latency. &
49\% lower queue length, similar throughput. &
1h (100 iters), $\leq\$15$ \\
\bottomrule
\end{tabularx}

\caption[Summary of project tasks]{Summary of project task objectives and corresponding SOTA publications, performance relative to SOTA and baseline solutions, and overall time/cost efficiency. 
Most tasks achieve near-SOTA performance within hours at modest cost, demonstrating the practicality of the ADRS approach.\protect\footnotemark}
\label{tab:project-summary}
\end{table*}

\begin{table*}[t]
\fontsize{7.4pt}{9pt}\selectfont
\setlength{\tabcolsep}{2.4pt}
\setlength{\extrarowheight}{1pt}
\centering
\begin{tabular}{l *{10}{r}}
\toprule
\textbf{Strategy} &
\textbf{Telemetry} &
\textbf{Cloudcast} $\downarrow$ &
\textbf{EPLB} &
\textbf{Prism} &
\textbf{LLM-SQL} &
\textbf{TXN} &
\textbf{CBL} $\downarrow$ &
\textbf{CBL-Multi} $\downarrow$ &
\textbf{MAS} &
\textbf{NS3} \\
\midrule
\rowcolor{gray!5}
\textbf{Human SOTA} &
0.8222 &
\bestnobold{626.24} &
0.127 &
21.892 &
0.6920 &
2724.8 &
101.68 &
92.332 &
\bestnobold{0.380} &
68.97 \\
\midrule
\multicolumn{11}{l}{\textbf{GPT-5}} \\
OE &
\textbf{0.930} $\pm$ 0.04 &
{926.9} $\pm$ 170.7 &
0.127 $\pm$ 0.00 &
26.23 $\pm$ 0.00 &
0.710 $\pm$ 0.01 &
\best{4238.6} $\pm$ 89.8 &
112.7 $\pm$ 4.2 &
81.47 $\pm$ 0.05 &
\textbf{0.333} $\pm$ 0.15 &
\textbf{92.24} $\pm$ 4.76 \\
GEPA &
0.916 $\pm$ 0.05 &
\textbf{689.9} $\pm$ 73.50 &
\best{0.134} $\pm$ 0.01 &
26.19 $\pm$ 0.07 &
\textbf{0.713} $\pm$ 0.00 &
3752.5 $\pm$ 204.4 &
\textbf{99.13} $\pm$ 1.49 &
\best{79.51} $\pm$ 1.50 &
0.290 $\pm$ 0.00 &
68.97 $\pm$ 0.00 \\
Shinka &
0.923 $\pm$ 0.04 &
954.8 $\pm$ 124.6 &
0.118 $\pm$ 0.01 &
\best{26.26} $\pm$ 0.00 &
0.712 $\pm$ 0.00 &
4090.0 $\pm$ 337.7 &
107.2 $\pm$ 1.7 &
79.84 $\pm$ 0.65 &
0.271 $\pm$ 0.11 &
89.50 $\pm$ 18.73 \\
\midrule
\multicolumn{11}{l}{\textbf{Gemini-3}} \\
OE &
\best{0.954} $\pm$ 0.01 &
\textbf{707.8} $\pm$ 40.1 &
\textbf{0.127} $\pm$ 0.00 &
26.24 $\pm$ 0.01 &
\best{0.729} $\pm$ 0.01 &
\textbf{4109.2} $\pm$ 253.9 &
107.4 $\pm$ 2.7 &
80.77 $\pm$ 0.71 &
0.188 $\pm$ 0.02 &
\best{115.2} $\pm$ 13.2 \\
GEPA &
0.850 $\pm$ 0.00 &
720.4 $\pm$ 46.2 &
\textbf{0.127} $\pm$ 0.00 &
26.16 $\pm$ 0.03 &
0.713 $\pm$ 0.00 &
3615.6 $\pm$ 481.2 &
\best{95.69} $\pm$ 0.50 &
81.45 $\pm$ 0.31 &
0.206 $\pm$ 0.01 &
74.47 $\pm$ 9.53 \\
Shinka &
0.918 $\pm$ 0.03 &
949.8 $\pm$ 73.4 &
0.120 $\pm$ 0.01 &
\textbf{26.25} $\pm$ 0.01 &
0.721 $\pm$ 0.00 &
3931.7 $\pm$ 343.4 &
105.0 $\pm$ 5.0 &
\textbf{80.26} $\pm$ 1.07 &
\textbf{0.243} $\pm$ 0.02 &
84.72 $\pm$ 7.70 \\
\bottomrule
\end{tabular}
\caption{Comparison of OpenEvolve (OE), GEPA, and Shinka across GPT-5 and Gemini-3.
We report mean $\pm$ standard deviation over three runs for each framework on downstream benchmarks. Higher the scores the better except for cases with $\downarrow$ label. }
\label{tab:oe-gepa-shinka-all}
\end{table*}

To rigorously evaluate the capability of \SYS/ in solving system performance problems, we investigate ten tasks across diverse sub-domains, including networking, databases, and core systems. We summarize our findings in Table~\ref{tab:project-summary}. Each case study follows a common schema that captures the problem setup, environment, evolutionary process, model usage, and final outcome, as detailed in Table~\ref{tab:expanded-schema} in the Appendix.

In this section, we present a subset of representative case studies selected to highlight insights to guide future research. These examples illustrate both the limitations of current frameworks and the best practices required to overcome them (detailed further in Section~\ref{sec:best-practices}). We evaluate three open-source ADRS frameworks (GEPA, OpenEvolve, and ShinkaEvolve) using their default configurations. To ensure a fair comparison, we equalize the generation budget by capping each run at 100 iterations using the GPT-5 and Gemini-3.0-Pro-Preview models. We repeat each experiment three times. The specific configs are provided in Appendix \ref{sec:config_files}.

We present the aggregate performance results across all \SYS/ use cases in Table \ref{tab:oe-gepa-shinka-all}. OpenEvolve achieved the highest success rate, delivering the best solution in 9 out of 20 total cases, followed by ShinkaEvolve with eight and GEPA with six (noting a tie between OpenEvolve and GEPA on the EPLB task with Gemini-3). When analyzing model sensitivity, we observe distinct behaviors: GEPA performs significantly better with GPT-5 (four top results) than with Gemini-3 (two top results), whereas ShinkaEvolve exhibits the inverse, favoring Gemini-3 (five top results) over GPT-5 (three top results). In contrast, OpenEvolve demonstrates the greatest versatility, maintaining consistent performance across both foundation models (four top results with GPT-5 and five with Gemini-3).

The five case studies cover distributed systems, databases, and LLM systems, summarized as follows:

\begin{itemize}

\item \emph{\textbf{CBL} - Optimizing Spot Instance Savings under Deadlines (Single Region):} Given a job with a deadline, the solution aims to maximize the use of cheaper spot instances in a public cloud without violating the deadline. \SYS/ improves the SOTA result by up to 35\% for a single region.

\item \emph{\textbf{CBL-Multi} - Optimizing Spot Instance Savings under Deadlines for Multi-Region:} An extension of the single-region spot instance scheduling problem to multiple regions, where the policy must also choose migration timing and region placement. \SYS/ achieves 17\% improvements over a strong baseline in a multi-region setting. 

\item \emph{\textbf{EPLB} - Optimizing Expert Placement in MoE Inference:} The solution seeks to balance the load across GPUs by mapping the expert replicas across GPUs. \SYS/ provides a fivefold improvement in the time it takes to rebalance experts compared with the best-known proprietary implementation.

\item \emph{\textbf{LLM-SQL} - Optimizing LLM Inference for SQL Queries:} The solution to this problem reorders rows and columns in a table to maximize the hit rate in a KV cache when performing LLM inference. \SYS/ achieves a similar hit rate to SOTA, while reducing the running time of the reordering algorithm by 3$\times$.

\item \emph{\textbf{TXN} - Optimizing Transaction Scheduling:} The solution aims to reorder transactions to minimize conflicts and hence improve the makespan and throughput. \SYS/ ``rediscovers'' the SOTA solution for the online case and improves a strong baseline by 34\% for the offline case, for which we are not aware of any published solution.

\end{itemize}

\footnotetext{Reported “time” reflects the automated solution-iteration phase 
(model–evaluator loop). It excludes researcher effort related to problem specification 
(e.g., refining prompts, evaluators, or experimental setup), which typically ranges from 
a few hours to a few days—still orders of magnitude less than manually developing a 
solution. Because this effort is difficult to measure consistently, we do not report it here.}

\subsection{Case Study \#1: Can't Be Late (CBL) [NSDI `24]} 
\label{sec:spot-instances}
\label{sec:cant-be-late}
Our first case study focuses on reducing the cost of deadline-driven jobs by exploiting cheaper but unreliable spot instances in the cloud. 
This problem was studied in an \textit{NSDI `24} outstanding paper~\cite{wu2024can}, which introduced the current state-of-the-art policy.
We report the performance of the best framework (GEPA) and show how it discovers an algorithm that improves the cost savings compared to the human SOTA by an average of 6\%, with per-workload improvements of up to 35\%.

\textbf{Problem setup.} 
The task is to minimize the cost of running deadline-aware jobs on a single node by using spot instances in one cloud region, while ensuring jobs still meets their deadlines. 
Spot instances are typically 60\% to 90\% cheaper than on-demand instances, but they may not always be available and can be preempted at any time. Each preemption incurs a changeover delay due to the setup time on a new instance. 

\textbf{Objective and constraints.} We evaluate the average cost savings across a range of workload traces. We require that all deadlines be met for an algorithm to be considered valid.

\textbf{Initial program and baselines.}  
The initial program is the \emph{greedy policy} from the original paper~\cite{wu2024can}, which uses spot instances until preemption risks missing a deadline. 
We compare against the \emph{Uniform Progress} algorithm, the paper's state-of-the-art solution, which tracks expected progress and switches between spot and on-demand instances based on whether it is ahead of or behind schedule.

\textbf{Solution generator and selector.} We run GEPA with Gemini-3.0 for 100 iterations. The search completes in about five hours and costs less than \$30.

\textbf{Evaluator.} We use the simulator from the NSDI paper and use configurations covering different job fractions, changeover delays, regions, and accelerator types. 
For each configuration, we sample 30\% of the traces as a feedback subset used during GEPA search to reduce overfitting to specific traces.
We report final results on the full evaluation set.

The evaluator checks syntax and interface compliance and tests valid solutions on sampled traces. 
It reports average cost savings over the Uniform Progress baseline and per-configuration statistics (mean, deviation, count). Trace features, such as availability and average spot duration, are also included to provide richer context.


\textbf{GEPA results.} 
The best policy achieves 6\% higher average cost savings than Uniform Progress (and 20\% over the greedy policy) while meeting all deadlines.
Per-trace improvements reach up to 35\% compared to Uniform Progress.


\begin{figure}[t]
\caption{Side-by-side comparison of the initial Uniform Progress policy and the evolved adaptive strategy. Key innovations in the evolved policy are highlighted.}
\label{fig:cant_be_late_evolution_comparison}
\centering

\begin{subfigure}[t]{0.48\textwidth}
\begin{minted}[
    frame=lines,
    fontsize=\scriptsize,
    linenos,
    breaklines=true
]{python}
def UniformProgress(has_spot, state, env, task):
    # Uniform progress assumption
    rate = task.duration / task.deadline
    expected = env.elapsed_time * rate
    actual = task.progress_made
    # Behind schedule: must run now
    if actual < expected:
        return SPOT if has_spot else ON_DEMAND
    # Simple hysteresis for on-demand
    if state == ON_DEMAND:
        buffer = expected + 2 * task.overhead
        if actual < buffer:
            return ON_DEMAND
    # Default preference
    return SPOT if has_spot else NONE
\end{minted}
\caption{Before: Uniform Progress Policy}
\label{fig:uniform-progress-code}
\end{subfigure}
\hfill
\begin{subfigure}[t]{0.48\textwidth}
\begin{minted}[
    frame=lines,
    fontsize=\scriptsize,
    linenos,
    breaklines=true,
    highlightlines={3-6,9-15,19-26,29-33,36-41},
    highlightcolor=blue!10
]{python}
def AdaptiveStrategy(has_spot, state, env, task):
    # Track spot availability history
    self.history.append(has_spot)
    avail = avg(self.history)
    trend = recent_avg(self.history) - avail
    # Stability via run-length statistics
    runs = extract_runs(self.history)
    stable = percentile(runs, 25)
    recover = percentile(runs, 50)
    cur_run = trailing_run(self.history)
    # Slack = remaining time - required time
    slack = time_remaining(env) - time_needed(task)
    # No slack: force ON_DEMAND
    if slack <= 0:
        return ON_DEMAND
    # Seal ON_DEMAND under tight slack
    if slack <= lock_margin(stable):
        self.sealed = True
    if self.sealed:
        if has_spot and cur_run >= recover:
            if slack > unlock_margin(recover):
                self.sealed = False
        if self.sealed:
            return ON_DEMAND
    # Anti-flapping dwell
    if state == ON_DEMAND and self.od_dwell < min_dwell():
        return ON_DEMAND
    # Buffer zone
    if slack <= retry_buffer():
        return SPOT if has_spot else ON_DEMAND
    # Safe zone with trend-aware switching
    if has_spot:
        if state == ON_DEMAND:
            thresh = switch_threshold(avail, trend)
            return SPOT if max(runs) >= thresh else ON_DEMAND
        return SPOT
    if state == ON_DEMAND:
        return ON_DEMAND
    # Adaptive waiting
    wait = slack * wait_factor(avail, trend)
    return NONE if self.wait_time < wait else ON_DEMAND
\end{minted}
\caption{After: Evolved Adaptive Policy}
\label{fig:adaptive-evolved-code}
\end{subfigure}

\end{figure}

As shown in Figure~\ref{fig:adaptive-evolved-code}, the evolved policy fundamentally differs from Uniform Progress shown in Figure~\ref{fig:uniform-progress-code}.
While Uniform Progress follows a fixed formula to maintain steady progress, the evolved policy tracks spot availability patterns and detects trends in availability over time (lines 3-6).
It estimates spot stability using percentiles of historical run lengths, making the thresholds robust to outliers (lines 7-10).
When the deadline approaches, the policy switches to on-demand instances and remains locked in that mode until spot availability is consistently stable (lines 16–24). To avoid repeatedly switching between instance types, the policy waits a minimum period after moving to on-demand before considering a switch back (lines 25–27). When there is sufficient time before the deadline, the policy factors in availability trends, returning to spot instances more readily when availability is high and improving (lines 32–36).

The main limitation of Uniform Progress is its inflexibility: it must use every available spot instance regardless of how short-lived it might be, leading to frequent switches with little progress. The evolved policy avoids this through adaptive waiting (lines 39–41): it waits longer when spot availability is high and improving, but switches to on-demand proactively when availability drops or begins declining.

\textbf{Evolution process.} The search explores the policy space through iterative refinement. Early candidates learn to track spot availability using a sliding window and compute the remaining time buffer. Subsequent candidates introduce a locking mechanism that switches to on-demand when the deadline approaches and only returns to spot instances after observing consistent availability. Later candidates refine stability estimation by switching from maximum run length to percentile-based thresholds, making the policy more robust to outliers. The final policy adds trend detection and adaptive waiting: it adjusts switching thresholds and wait times based on whether spot availability is improving or declining.

\subsection{Case Study \#2: \\Multi-Region Can't Be Late}
\label{sec:cant-be-late-multiple-regions}

In this section, we show that OpenEvolve develops an algorithm on an expanded multi-region setting, where no prior policy has been published. In this setting, OpenEvolve discovers an algorithm that outperforms a hand-tuned baseline by 17\%.

\textbf{Problem setup.} The original Uniform Progress assumes one region with uniform spot prices. In practice, spot prices and availability differ across regions (\cite{10.1145/3689031.3717459}), so a policy must decide when to switch spot and on-demand, which region to use, and when to migrate jobs. We use OpenEvolve to explore this multi-region space and derive an better policy.

\textbf{Objective and constraints.} We evaluate the total cost in a multi-region setup, accounting for spot and on-demand instances and migration costs. A policy is valid only if all job deadlines are met.

\textbf{Initial program and baselines.} As no baseline exists, we adopt a Uniform Progress variant that first assigns spot instances locally and, if none are available, move to other regions in a round-robin manner.


\textbf{Solution generator and selector.}
We run OpenEvolve for 100 iterations with GPT-5, using the same island setup as before. The evaluator then reports both overall and per-trace scores, similar to Section~\ref{sec:cant-be-late}.

\textbf{Evaluator.} For the multi-region setting, we extend the Uniform Progress simulator and evaluate cost savings on 106 traces.

\textbf{OpenEvolve results.}
The final policy achieves 17\% cost savings, on average, compared to the multi-region Uniform Progress baseline. It balances cost efficiency with deadline guarantees using a simple principle: when a job is not urgent (i.e., not at risk of missing its deadline), it explores additional regions to seek lower-cost spot capacity; if a job is urgent, it prioritizes immediate progress, selecting spot instances when available or falling back to on-demand. This adaptive logic enables opportunistic exploration under slack conditions while ensuring reliability when deadlines are at risk, effectively managing the trade-off between exploration and guaranteed progress in a multi-region environment. In addition, the policy leverages a dynamic view of regional capacity to opportunistically migrate when conditions are favorable.

\textbf{Evolution process.} The search process demonstrates iterative improvement of the deadline monitoring mechanisms from multi-region scheduling policies. 
Initial strategies implement basic progress tracking by comparing task completion against elapsed time.
The system discovers key insights through failure analysis.
At iteration 7, the system introduces region caching and urgency calculation. Iteration 5-12 attempts with aggressive cost reduction initially show promise, but ultimately fail when accumulated delays cannot be recovered within deadline constraints. These failures guide the search toward more balanced approaches.
The final evolved strategy at iteration 63 implements a two-stage urgency detection system. Rather than applying uniform resource allocation rules, it combines schedule-based progress monitoring with direct deadline pressure analysis. This design enables adaptive behavior: immediate allocation of on-demand instances when deadlines are at risk, while maintaining intelligent region exploration when deadline permits. The insight is the separation of deadline assessment from resource provisioning decisions, enabling adaptive region selection.






\subsection{Case Study \#3:  Expert Placement in MoE Inference (EPLB)} 
\label{sec:EPLB}

In this section, we use ADRS framworks to design algorithms to balance inference load across multiple GPUs in Mixture-of-Experts (MoE) architectures. In our evaluation, ShinkaEvolve discovers an implementation that is 13.0$\times$  faster than a proprietary, frontier-lab baseline while achieving a similar load-balance factor.
 
\textbf{Problem setup.} The basic Expert Parallelism Load Balancing (EPLB) algorithm operates in three stages: (i) distributing expert groups across nodes to balance the load, (ii) creating replicas for hot (popular) experts, and (iii) assigning these replicas to  GPUs to maximize load balance. The problem is to determine the replica count per expert and map expert replicas to GPUs to maximize load balance for a given query workload, MoE model, and GPU set.

\textbf{Objective and constraints.} 
Our goal is twofold: maximize load balance (i.e., the ratio of average to maximum tokens generated per GPU) and minimize the runtime of the algorithm when the load distribution shifts.

\textbf{Initial program and baselines.} The initial program is the open-source implementation from DeepSeek~\cite{deepseek-eplb}, which performs expert placement using a greedy bin-packing. Specifically, it sorts experts by load and assigns them iteratively to the least-loaded feasible GPU. While simple, its Python-based linear search using a for-loop is slow, averaging 693 ms for a 0.66 balance factor. 
We also include a non-public reference implementation from a frontier lab as a baseline. This implementation avoids explicit iteration and reduces the runtime to 19.5 ms while achieving the same balance factor as the open-source algorithm.

\textbf{Solution generator and selector.} We employ the same setup used for all use cases (Section 4). For each framework, evolution takes roughly five hours and costs under \$15.

\textbf{Evaluator.} 
Our simulator models a distributed GPU inference engine for MoE models. The simulator is implemented in 168 lines of PyTorch code. Our evaluation trace models load changes over the ShareGPT and GSM8K datasets~\cite{cobbe2021training, vicuna2023}. 

The evaluator's output metrics are: (a) the balance factor and (b) the time it takes to rearrange the expert replicas during the load changes. We compute the combined score as the equally weighted average of the load balance factor and a normalized lower-is-better runtime term.

\begin{figure}[t]
\centering
\caption{Side-by-side comparison of the initial greedy policy and final evolved heuristic. Key innovations in the evolved policy are highlighted.}
\label{fig:evolution_comparison}

\begin{subfigure}[t]{0.48\textwidth}
\begin{minted}[
    frame=lines,
    fontsize=\scriptsize,
    linenos,
    breaklines=true
]{python}
def InitialStrategy(...):
    ...
    for item in sorted(items, reverse=True):
        # Greedily choose least-loaded pack
        # Plain iterative assignment
        available = filter_nonfull(packs)
        target = min(available)
        target.add(item)
    ...
\end{minted}
\caption{Before: Initial Program}
\label{fig:eplb-initial-code}
\end{subfigure}
\hfill
\begin{subfigure}[t]{0.48\textwidth}
\begin{minted}[
    frame=lines,
    fontsize=\scriptsize,
    linenos,
    breaklines=true,
    highlightlines={12-18},
    highlightcolor=blue!10
]{python}
def EvolvedStrategy(...):
    ...
    # Items are pre-sorted by weight
    idx = arange(num_items)

    block  = idx // num_packs
    offset = idx % num_packs

    # Vectorized snake assignment
    # even blocks: 0 -> P-1
    # odd blocks:  P-1 -> 0
    pack_id = where(
        block % 2 == 0,
        offset,
        num_packs - 1 - offset
    )

    assign(items, pack_id)
    ...
\end{minted}
\caption{After: Evolved Policy}
\label{fig:eplb-evolved-code}
\end{subfigure}
\end{figure}

\textbf{ShinkaEvolve results.} 
The evolved algorithm 
replaces the linear Python for-loop with vectorized tensor operations: instead of explicit greedy bin packing, it constructs a tensor of expert indices and assigns experts via a zigzag (“snake”) pattern across GPUs. This assignment alternates direction across contiguous blocks of experts, interleaving higher- and lower-load experts across GPU slots.
The resulting algorithm matches the load balance factor of the other baselines while reducing runtime to just 1.51 ms, yielding a 13.0$\times$ speedup over the proprietary implementation. 

\textbf{Evolution process.}
ShinkaEvolve's evolution trajectory can be characterized by two major steps in improving runtime: first, replacing Python for-loops with PyTorch tensor operations, and second, discovering the zigzag placement pattern (Figure~\ref{fig:eplb-evolved-code}). Interestingly, the initial introduction of the zigzag pattern did not yield immediate gains—the balance factor sometimes worsened, and rearrangement costs fluctuated. The breakthrough came later, when the evolution process learned to systematically reuse the zigzag partitioning heuristic across multiple stages of EPLB, rather than only in the initial group distribution. 
At this point, both runtime and stability improved substantially in our trace.”

The expert replication stage, by contrast, remained the most unstable throughout evolution. The system oscillated between strategies such as copying the least-used experts, overloading popular ones, or attempting proportional spreads. These experiments rarely improved the score, and ultimately the intuitive rule of replicating only overloaded experts prevailed. Consequently, many iterations were unproductive, with the main speed improvements coming from global reorganization logic that exploited PyTorch’s batched operations and the zigzag layout.

In summary, ShinkaEvolve rediscovered and exploited a tensorized zigzag partitioning scheme, yielding an evolved EPLB algorithm that achieves a 13.0$\times$ speedup while maintaining the same balance factor.

\subsection{Case Study \#4: LLM Inference on SQL Queries (LLM-SQL) [MLSys `25]} 
\label{subsec:llm-sql}
This research problem~\cite{liu2024optimizing} arises in relational analytics, where SQL queries invoke LLMs over entire tables, with each row triggering a separate LLM inference operation. At scale, this is prohibitively expensive. The state-of-the-art solution mitigates cost by reordering rows and fields to maximize prefix KV cache reuse. 
Using OpenEvolve, we evolve such a reordering policy, achieving similar hit rates while delivering a 3$\times$ runtime speedup.

\begin{figure}[h!]
\centering
\caption{Side-by-side comparison of the greedy recursive grouping (GGR) and the evolved prefix-aware reordering policy. Key innovations in the evolved algorithm are highlighted.}
\label{fig:reorder_comparison}

\begin{subfigure}[t]{0.48\textwidth}
\begin{minted}[
frame=lines,
fontsize=\scriptsize,
linenos,
breaklines=true
]{python}
def GGR(df):
    # 1. Compute value counts for all cells
    counts = Counter(df.stack())
    val_len = {v: len(str(v))**2 for v in counts}

    # 2. Pick value maximizing len^2 * (count-1)
    v_star = argmax_v [ val_len[v] * (counts[v]-1) ]
    if v_star is None: return fixed_reorder(df)

    # 3. Split rows with/without v_star
    G = rows with v_star
    R = rows without v_star

    # 4. Reorder columns in G (v_star front, deps after)
    for row in G:
        cols_with_val = [c for c in df.columns if row[c]==v_star]
        reordered = cols_with_val + (others)
        row = row[reordered]

    # 5. Recurse on G remainder and R
    G = QuickGreedy(G remainder)
    R = QuickGreedy(R)
    return concat(G,R)
\end{minted}
\caption{GGR baseline.}
\label{fig:quick_greedy_code}
\end{subfigure}\hfill%

\begin{subfigure}[t]{0.48\textwidth}
\begin{minted}[
frame=lines,
fontsize=\scriptsize,
linenos,
breaklines=true,
highlightlines={4-7,11-16,18-23,27-31},
highlightcolor=blue!10
]{python}
def EvolvedPolicy(df):
    # 1. Precompute value statistics ONCE
    counts  = Counter(df.values.ravel())
    val_len = {v: len(str(v))**2 for v in counts}

    BASE = 5000   # size threshold to cap recursion depth

    # 2. Select best grouping value (GGR score)
    score = lambda v: val_len[v] * (counts[v] - 1)
    v_star = max(counts, key=score, default=None)
    if v_star is None or counts[v_star] <= 1:
        return FixedReorder(df)

    # 3. Split rows by whether they contain v_star
    grouped   = df[df.eq(v_star).any(axis=1)]
    remaining = df[~df.eq(v_star).any(axis=1)]

    # 4. Column recursion: pull v_star forward (stable)
    grouped = StableColumnReorder(grouped, v_star)

    # 5. Recursive descent with early stopping
    if len(df) > BASE:
        top, bot = split(df)
        return concat(EvolvedPolicy(top),
                      EvolvedPolicy(bot))

    # 6. Recurse on remainder and concatenate
    if not remaining.empty:
        return concat(
            grouped,
            EvolvedPolicy(remaining)
        )

    return grouped
\end{minted}
\caption{Evolved prefix-aware policy.}
\label{fig:evolved_policy_code}
\end{subfigure}

%






\end{figure}

\textbf{Problem setup.} To minimize inference time and cost, we aim to maximize the prefix cache hit rate (PHR) by reordering both rows and fields in the table before performing inference. This problem is combinatorial: for a table with $n$ rows and $m$ fields, there are $n! \times (m!^n)$ possible orderings, making a naive brute-force search infeasible. Thus, the goal is to design a reordering algorithm that achieves high PHR while keeping its runtime small relative to the overall inference time.

\textbf{Objective and constraints.} Our objective is to maximize the prefix hit rate (PHR) while keeping the runtime of the reordering algorithm low. Since PHR measures the fraction of token prefixes shared across consecutive rows, and serves as a proxy for inference cost and latency. 

\textbf{Initial program and baselines.} 
As initial program, we use the greedy recursive group algorithm (GGR)~\cite{liu2024optimizing}, a heuristic that recursively groups rows by common field values and reorders fields using schema statistics with early stopping. This approach approximates the optimal reordering algorithm while running more efficiently. For comparison, we also include a simple baseline: the table in its original ordering, i.e., the default row/field order of the input table. This program is implemented in Pandas~\cite{mckinney2010data}, an open-source Python library for data manipulation and analysis.


\textbf{Solution generator and selector.} We configure OpenEvolve with three islands, and we use an LLM ensemble of 80\% OpenAI o3 and 20\% Gemini 2.5 Pro, and we run it for 100 iterations. The entire evolution takes about one hour and costs less than \$7.

\textbf{Evaluator.} We leverage the publicly available simulator from the paper that measures prefix cache hit rate (PHR) given dataset table. 
The simulator is written in Python (200 LOC) and evaluates a benchmark of representative LLM queries across five recommendation datasets (e.g., movies~\cite{rotten-tomatoes-movies-dataset}, beer~\cite{ratebeer}, BIRD~\cite{bird}, PDMX~\cite{pdmx}, products~\cite{amazon-product-review-dataset}).

The evaluator reports four key metrics. First, it reports a combine score $ = 0.95 \times \text{PHR} + 0.05 \times \frac{1}{1 + \text{runtime}}$, defined as the equally weighted average of PHR and algorithm runtime across datasets, where higher PHR and lower runtime yield a higher score. To preserve query semantics, the reordering algorithm must not alter which rows or fields are included, only their order. Second, the evaluator records a binary flag indicating whether the candidate program executes end-to-end. Third, it reports the detailed prefix hit rate for each dataset. Finally, it measures the total runtime of the policy. The combined score serves as the optimization objective during evolution.


\textbf{OpenEvolve results.} The program synthesized by OpenEvolve attains a comparable average PHR to the GGR algorithm while reducing end-to-end runtime by approximately $3\times$, resulting in a higher overall combined score. The greedy baseline incurs substantial overhead due to repeated value-count recomputation and deep recursive traversal over the table. In contrast, the evolved implementation incorporates several structural optimizations. First, it avoids redundant scans by maintaining a cached global value-frequency map and lazily updating value-length statistics, eliminating repeated full-table passes. Second, it replaces expensive Pandas-based indexing and grouping operations with direct NumPy and Python-level array manipulations, reducing the dominant computation to simple $O(N_{\text{rows}} \times N_{\text{cols}})$ loops. Finally, rather than globally re-sorting the table at each step, the algorithm applies a localized, prefix-aware reordering heuristic that prioritizes continuity with the previous row while weighting values by squared string length. These optimizations, illustrated in Figure~\ref{fig:reorder_comparison}, jointly explain the substantial runtime improvements without sacrificing PHR.


\textbf{Evolution process.} 
The search begins with the published GGR algorithm, which achieves a good Prefix Hit Rate (PHR) but suffers from repeated counter operations and deep recursion. Early in the search, by iteration 32, OpenEvolve discovers a faster heuristic that orders columns by (frequency $\times$ squared length) instead of using recursive splitting. This change eliminates redundant value counting and improves runtime, though at the cost of some grouping accuracy.

Later, by iteration 72, the heuristic is refined. It now uses normalized weights (frequency ratio $\times$ squared length) and limits multi-key sorting to only the most informative columns, which improves hit rates while maintaining efficiency.

By iteration 97, the final program strikes an optimal balance between speed and accuracy. It raises the recursion base threshold, reuses cached counts and string lengths, reintroduces selective recursion for rows, and incorporates a NumPy-based reordering to replace costly pandas lookups.

Overall, the evolution progresses from early runtime improvements to mid-stage refinements that recover accuracy, culminating in a final design that integrates both for the best overall score.

\begin{figure}[h!]
\centering
\caption{Side-by-side comparison of the evolved constant-time greedy policy (SMF) and the offline evolved policy. Variable names are abbreviated for brevity without losing logic.}
\label{fig:schedule_comparison_numbered}

\begin{subfigure}[t]{0.48\textwidth}
\begin{minted}[
frame=lines,
fontsize=\scriptsize,
linenos,
breaklines=true,
tabsize=2
]{python}
# --- 1. Candidate sampling --- #
K = 8
rng = np.random.default_rng()
while rem:
  # --- 2. Draw candidates --- #
  # Handle tail end of queue
  if len(rem) <= K: cands = rem
  else: cands = rng.choice(rem, K, replace=False)
  # --- 3. Evaluate incremental cost --- #
  best_cand = None
  best_add = math.inf
  best_times = (0, 0)
  for cand in cands:
    delta, t0, t1 = calc_cost(cand, state)   
    if delta < best_add:
      best_add = delta
      best_cand = cand
      best_times = (t0, t1)
  # --- 4. Commit update global state --- #
  schedule.append(best_cand)
  rem.remove(best_cand)
  # Update locks/occupancy maps
  update_state(best_cand, best_times)
  total_cost += best_add
return total_cost, schedule
\end{minted}
\vspace{-1em}
\caption{SMF / greedy policy.}
\label{fig:smf}
\end{subfigure}\hfill%
\begin{subfigure}[t]{0.48\textwidth}
\begin{minted}[
frame=lines,
fontsize=\scriptsize,
linenos,
breaklines=true,
tabsize=2,
highlightlines={9-13, 18-30, 33-36,40-43},
highlightcolor=blue!10
]{python}
def get_best_schedule(workload, n_seqs):
  n = workload.num_txns
  # --- 1. Cost Cache --- #
  memo = {}
  def get_cost(s):
    return memo.setdefault(tuple(s), workload.eval(s))
    
  # --- 2. Pairwise Scoring (O(N^2)) --- #
  score = [0] * n
  for i in range(n):
    for j in range(i + 1, n):
      d = get_cost([j, i]) - get_cost([i, j])
      score[i] += d; score[j] -= d
  best_c, best_s = float("inf"), None
  
  # --- 3. Multi-Start Strategy --- #
  for k in range(max(6, n_seqs)):
    mode = k % 3
    
    if mode == 0:   # A. Greedy w/ Random Sampling
      seq = [random.choice(range(n))]
      rem = list(set(range(n)) - set(seq))
      while rem:
        batch = random.sample(rem, min(len(rem), 12))
        # ... (evaluate best insertion pos) ...
        seq.insert(best_pos, best_cand)
    elif mode == 1: # B. Beam Search
      seq = beam_search_helper(n, get_cost)
    else:           # C. Borda Count Sort
      seq = sorted(range(n), key=lambda x: -score[x])
      
    # --- 4. Deterministic Polish --- #
    improved = True
    
    while improved:
      # ... (scan swaps/moves, update seq) ...
      improved = try_local_moves(seq)
    # --- 5. Final Refinement --- #
    # Perturb + final deterministic polish
    seq = escape_local_optima(seq, get_cost)

    if (c := get_cost(seq)) < best_c:
      best_c, best_s = c, seq

  return best_c, best_s
\end{minted}
\vspace{-1em}
\caption{Offline evolved policy.}
\label{fig:txn_offline}
\end{subfigure}

\end{figure}

\begin{table*}[t]
\centering
\scriptsize
\setlength{\tabcolsep}{4pt}
\renewcommand{\arraystretch}{1.3}

\begin{tabularx}{\textwidth}{@{}p{3cm} p{3.5cm} p{3.5cm} X@{}}
\toprule
\textbf{Problem} & \textbf{Technique} & \textbf{Domain Origin} & \textbf{How it helps?} \\
\midrule

\textbf{EPLB} \newline (GEPA, GPT~5) &
\textbf{Hamilton's Apportionment} &
\textbf{Political Science} \newline (US Congress History) &
\textbf{Fast-Path Allocation:} Provides a baseline for assigning integer replica counts using the ``Largest Remainder Method.'' It is used for uniform workloads to avoid expensive search iterations. \\
\midrule

\textbf{Transaction Scheduling} \newline (OpenEvolve, GPT-5) &
\textbf{Condorcet / Borda Count} &
\textbf{Economics} \newline (Social Choice Theory) &
\textbf{Preference Aggregation:} Constructs a matrix where every transaction ``votes'' on its precedence. The score is the aggregated cost savings of essentially running an election to determine the ``consensus'' order. \\
\midrule

\textbf{Telemetry Repair} \newline (OpenEvolve, Gemini-3) &
\textbf{Kirchhoff's Current Law} \newline (KCL) &
\textbf{Electrical Engineering} \newline (Fluid Dynamics) &
\textbf{Constraint Satisfaction:} Enforces the physical law of ``flow conservation'' to validate noisy telemetry counters. Deviations indicate data errors. \\
\midrule



\textbf{Can’t Be Late Multi-Region} \newline (GEPA, GPT-5) &
\textbf{Multi-Armed Bandit with UCB-style Scoring} &
\textbf{Artificial Intelligence} \newline (Reinforcement Learning) &
\textbf{Explore-Exploit Region Selection:} Maintains per-region success/failure counters (succ, fail) to compute a ``success rate'' score. Unknown regions are prioritized for exploration, while high-success regions are exploited. This mirrors the Upper Confidence Bound (UCB) algorithm for balancing exploration vs.\ exploitation in slot machine selection. \\
\midrule

\textbf{Datacenter TCP Congestion Control} \newline (OpenEvolve, Gemini-3) &
\textbf{Gradient-based Power Control} &
\textbf{Physics / Electrical Engineering} \newline (Electrical Circuits \& Power) &
\textbf{Congestion Anticipation:} Adapts the physical concept of ``Power'' ($P = V \cdot I$) to networking, treating throughput as Current and delay as Voltage. It computes the RTT gradient ($\frac{dRTT}{dt}$) to measure the ``velocity'' of queue growth, allowing the system to react to the energy of congestion bursts before queues overflow. \\

\bottomrule
\end{tabularx}

\caption{Cross-domain techniques discovered by ADRS frameworks on different problems.
The “Problem’’ column denotes the framework and model that generated the best-performing program for the given systems problem.
The table is not exhaustive and presents only one representative technique per problem.}
\label{tab:cross-domain-techniques}
\end{table*}

\subsection{Case Study \#5: Transaction Scheduling (TXN) [VLDB `24]}
\label{sec:transaction-scheduling}
This research problem~\cite{cheng2024towards} aims to find efficient schedules to reduce conflicts for transactional workloads. 
Optimizing execution order of transactions can significantly improve throughput by minimizing overall execution time. We apply OpenEvolve and find it is unable to find a solution that outperforms the state-of-the-art policy under the original online setting constraints. However, OpenEvolve discovers a superior scheduling algorithm in the offline setting, demonstrating the utility of ADRS frameworks for rapidly exploring problem variations.

\textbf{Problem setup.} Conflicts on shared data cause performance bottlenecks in many transactional workloads~\cite{taobench}. One approach to minimize conflicts is to carefully schedule the transactions. The problem we aim to solve is: given a set of transactions, find a schedule that minimizes the conflicts and improves the throughput. 

\textbf{Objective and constraints.} Maximizing throughput in this setting is equivalent to minimizing the schedule \textit{makespan}, i.e., the total time to execute all transactions. We consider both the online and the offline settings. In the online setting, we assume that the transaction order is fixed once the schedule is determined (i.e., committed transactions cannot be rollbacked). We constrain the scheduling algorithm to O($n$) runtime (where $n$ is the number of transactions to be scheduled) and assume the read/write operations are not known apriori (only hot keys can be predicted). We also consider the offline scheduling problem, which is relevant to deterministic databases~\cite{ren2019slog, calvin2012} that schedule batches of transactions, a setting with no previously known results.

\textbf{Solution generator and selector.} 
We configure OpenEvolve to run with three islands and use OpenAI's GPT-5 for both problem settings. We run for 100 iterations, which takes less than two hours and costs less than \$20. 

\textbf{Initial program and baselines.} We use a random scheduler for the initial program. We compare against a number of transactional scheduling algorithms, including the SOTA algorithm, Shortest Makespan First (SMF), which greedily chooses transactions to schedule.

\textbf{Evaluator.} We use the Python simulator from the SMF paper~\cite{cheng2024towards}, which assumes that each operation takes one unit of time. The simulator calculates the makespan of a given transaction schedule and also provides statistical bounds on the makespan of the schedule for a given workload. We measure total makespan over five traces from the OLTP benchmarks used in the original paper (Epinions, SmallBank, TPC-C, TAOBench, YCSB) with 500 transactions each. We use the random-scheme baseline as the initial program. 

\begin{table*}[t]
    \centering
    \includegraphics[width=\linewidth]{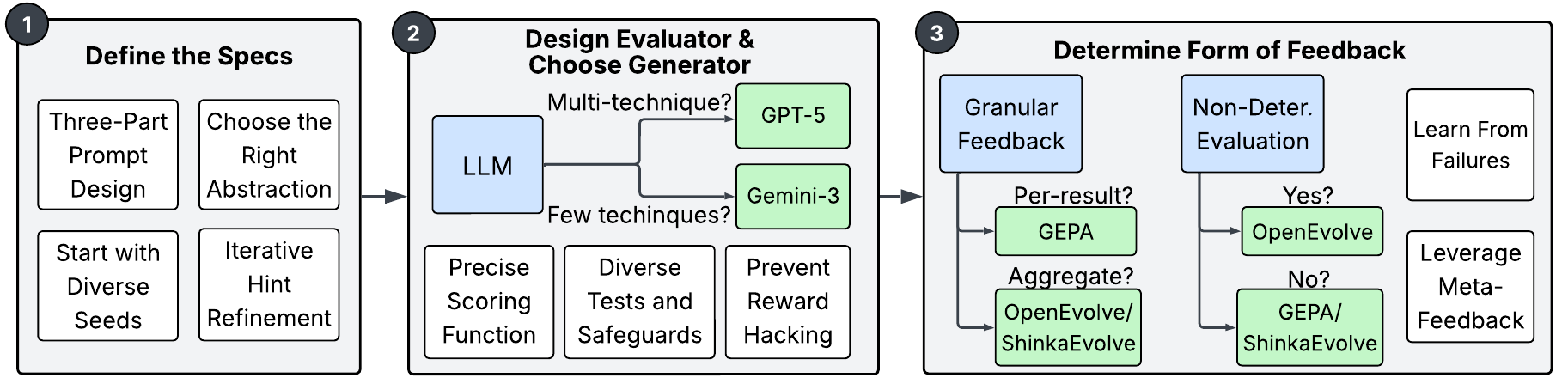}
    \captionof{figure}{Workflow diagram that illustrates how these insights integrate into the end-to-end ADRS process.\protect\footnotemark}
    \label{fig:ADRS_workflow}
\end{table*}

\begin{table*}[t]
\centering
\small
\setlength{\tabcolsep}{6pt}
\renewcommand{\arraystretch}{1.15}

\begin{minipage}[t]{\textwidth}
\centering
\begin{tabularx}{\linewidth}{p{6.85cm} X}
\toprule
\textbf{ADRS Axis} & \textbf{Actionable Insights} \\
\midrule

\textbf{Specifications: \textit{"Less is More and More is Less"}} &
\vspace{-1.1em}
\begin{itemize}[leftmargin=*]
    \item Three-part prompt with problem, evaluation criteria, and context.
    \item Choose the right abstraction to match the optimization goal.
    \item Start evolution with diverse, domain-specialized seeds.
    \item Iteratively refine hints to balance explore-exploit.
\end{itemize}
\\
\midrule

\textbf{Evaluation: \textit{"Solution is Only as Good as Evaluator"}} &
\vspace{-1.1em}
\begin{itemize}[leftmargin=*]
    \item Enforce diverse test sets and safeguards.
    \item Prevent reward hacking by restricting edits.
    \item Use precise scoring function that is smooth and deterministic.
    \item Select LLM based on solution composition.
\end{itemize}
\\
\midrule

\textbf{Feedback: \textit{"The Devil is in the Details"}} &
\vspace{-1.1em}
\begin{itemize}[leftmargin=*]
    \item Calibrate feedback granularity to provide actionable guidance.
    \item Learn from failures to gain insights for evolution.
    \item Meta-feedback helps guide search when accurate.
    \item Encode resilience for non-deterministic evaluators.
\end{itemize}
\\
\bottomrule
\end{tabularx}
\caption{Summary of actionable insights for ADRS, covering specification, evaluation, and feedback.}
\label{tab:adrs_summary_crisp}
\end{minipage}
\hfill
\end{table*}

\textbf{OpenEvolve results.} In the online setting, OpenEvolve's best policy matches SMF. OpenEvolve rediscovered this algorithm from a random baseline, likely due to training data contamination from the SMF paper. While this result is not as interesting, this confirms ADRS frameworks can reproduce state-of-the-art solutions. Surpassing SMF is difficult because transaction scheduling involves complex operation groups, dependencies, and correctness constraints (e.g., serializability). Prior work showed that complex heuristics (hill climbing, simulated annealing) did not perform better due to these constraints~\cite{cheng2024towards}. Instead, SMF's strategy of leveraging conflict costs provides a generalizable solution across workloads.

At a high level, SMF minimizes contention by separating transactions with high conflict costs. The incremental makespan increase of adding a transaction to the schedule accounts for all potential conflicts with the current ordering. Specifically, SMF starts with a random transaction. At each iteration (Figure~\ref{fig:smf}, lines 5), it selects the transaction that increases makespan the least among k sampled unscheduled requests (lines 8-22) and appends it to the schedule (lines 25-30). Ties are broken randomly. The algorithm runs in linear time $O(n \times k)$, where $n$ is the transaction count and $k$ is the sample size.

In the offline setting, OpenEvolve discovers a novel algorithm that reduces makespan by 60\% compared to SMF. This result reduces concerns about contamination since it is not a previously known solution. The evolved policy employs a multi-stage strategy. First, it computes pairwise conflict costs to assign a score to each transaction (Figure 11, lines 9-14). Second, it iterates through a multi-start loop, generating candidates via randomized greedy insertion, beam search, or score-based sorting (lines 18--32). Each candidate undergoes deterministic polishing to resolve local minima (lines 36--38) and a final perturbation step to escape optima (line 42). This algorithm extends the greedy intuition of SMF while maintaining $O(n^2)$ runtime. This result indicates that scheduling based on conflict costs is the effective approach for reducing makespan. Furthermore, OpenEvolve rapidly adapts the algorithm for altered constraints, showing it can assist researchers in customizing solutions for different settings (which typically require manual re-design).

\textbf{Evolution Process} 
In the online setting, OpenEvolve's search shifts from random schedules and length-only heuristics toward conflict-aware solutions. Recognizing that contention increases makespan, candidates explore write-count bucketing and greedy contention minimization. Common pitfalls included over-reliance on transaction length (which correlates poorly with contention) and expensive greedy selection that violated the $O(n)$ constraint. Early candidates often converge prematurely on single heuristics (e.g., length-first, write-first). The final policy generalizes these insights, incorporating conflict costs to match SMF.

For the offline setting, the evolution also shifts from random sampling to structured construction heuristics (e.g., beam search, pairwise scoring). The search eventually centers in two directions: (i) multi-start construction strategies that build schedules incrementally and (ii) guided local search that refines them. The search often finds that standard hill climbing plateaued quickly in local minima. To overcome this, the framework discovers the importance of pairwise precedence by analyzing whether executing transaction $i$ before $j$ is cheaper than the reverse. The best solution combines these insights: it employs a mix of initialization strategies (greedy insertion, beam search, and Borda sorting based on precedence scores) followed by a polishing phase that uses pairwise-guided swaps and block moves to escape local optima.

\subsection{Cross-Domain Analysis}
\label{appendix:cross-domain}

One of the most compelling capabilities of ADRS is its ability to generate solutions by drawing on techniques from different domains. While systems researchers typically specialize in a particular subfield (e.g., networking or databases), LLMs are trained on vast, diverse corpora that span the entire spectrum of human knowledge~\cite{brown2020languagemodelsfewshotlearners,chowdhery2022palmscalinglanguagemodeling}. This breadth allows ADRS to identify structural parallels between systems performance problems and those in other fields, revealing connections that may be overlooked by a human expert focused on a single domain. As hypothesized in Section~\ref{subsec:sys}, this cross-domain reasoning enables ADRS to import techniques from fields like political science, economics, and physics to solve modern systems challenges efficiently. Table~\ref{tab:cross-domain-techniques} shows representative cross-domain techniques discovered by ADRS frameworks across our case studies. For each problem, the table reports one example technique, its domain of origin, and how it is used in the generated solution. The listed techniques originate from areas such as political apportionment, voting systems, electrical engineering, combinatorial optimization, real-time scheduling, reinforcement learning, and congestion-control modeling. The table is not exhaustive for space and provides a single example technique per problem.

In \textbf{EPLB} (Section~\ref{sec:EPLB}), the core challenge is to assign integer replica counts to experts to balance load. This is a structurally identical problem to allocating congressional seats to states based on the voting population. GEPA, using GPT-5, ``rediscovered'' Hamilton’s Apportionment method (also known as the Largest Remainder Method) from political science. By applying this 18th-century algorithm, the system implemented a fast-path allocation strategy for uniform workloads that avoided expensive search iterations, contributing to the $13\times$ speedup over the baseline.

Similarly, in \textbf{TXN}  (Section~\ref{sec:transaction-scheduling}), OpenEvolve leverages concepts from social choice theory and economics to minimize conflicts. To determine the optimal execution order, the generated solution constructs a precedence matrix. where transactions effectively ``vote'' on their preferred ordering based on conflict costs, a technique mirroring the Condorcet or Borda Count methods used in voting systems. This approach allowed the scheduler to determine an order that minimizes global makespan, effectively translating a preference aggregation problem into a high-performance database schedule. 

\section{Best Practices}
\label{sec:best-practices}
Having demonstrated the potential of \SYS/ to surpass state-of-the-art solutions, we now focus on the most effective strategies for applying these frameworks to algorithm discovery. Drawing from extensive ablation studies across the evolutionary process, we consolidate our findings into a set of actionable best practices. We categorize these strategies across three critical axes, \emph{specifications}, \emph{evaluation}, and \emph{feedback} and summarize them in Table~\ref{tab:adrs_summary_crisp}. Additionally, we provide a workflow guide in Figure~\ref{fig:ADRS_workflow} for applying \SYS/ frameworks to new research problems. 

\subsection{Specs: Less is More and More is Less}
Rigorously defined specifications are a prerequisite for effective search. We find that ambiguity in prompts is a primary source of execution and algorithmic failures (e.g., missing API details). Effective prompts should be highly specific and structured, clearly defining three key areas:

\begin{itemize}[noitemsep,topsep=0pt,parsep=0pt,partopsep=0pt]
\item \emph{The problem:} the core problem definition.
\item \emph{The evaluation criteria:} optimization goals and correctness constraints.
\item \emph{The context:} essential background, such as required APIs. 
\end{itemize}

We recommend instantiating a concise, three-section template for each problem and polishing the draft with external LLMs (e.g., ChatGPT and Gemini) prior to evolution. In our experience, this has been crucial to reducing invalid generations and preventing evaluator hacking. 

\textbf{[Prompt Generator] High-level or Low-level abstractions?} 
Selecting the appropriate level of abstraction for evolution is crucial to balancing implementation convenience with genuine problem-solving. Unrestricted access to high-level external library APIs can lead to suboptimal optimizations. With \textbf{EPLB}, replacing custom operators with PyTorch primitives yielded only trivial speedups.
To encourage algorithmic advances, we can restrict API access to help the ADRS framework explore new strategies rather than relying on pre-built solutions. Conversely, providing high-level APIs can help the ADRS framework to focus on the desired logic to optimize. With \textbf{Cloudcast}, exposing simulator internals, boilerplate state management, or concurrency primitives in the prompt causes the LLM to waste iterations tuning irrelevant implementation details rather than making algorithmic improvements. The right abstraction is a minimal, higher-level interface containing only the target function signature (e.g., \texttt{search\_algorithm()}) and essential helper classes (e.g., \texttt{make\_nx\_graph()} to return cost profiles); this provides a focused starting point for more effective search. 

\footnotetext{Specific model recommendations (e.g., GPT-5 vs. Gemini-3.0) represent a snapshot based on ablation studies across our 10 use cases. These serve to illustrate design space trade-offs; we expect optimal generator choices to evolve alongside model capabilities.}



\takeaway{1}{Choose the right abstraction by carefully scoping available APIs to match the optimization goal.}

\textbf{[Prompt Generator] Basic or SOTA initial programs?}
The choice of base program shapes the trajectory of algorithm evolution. Buggy or weak baselines waste iterations on trivial fixes, while strong baselines can accelerate progress.
For example, in \textbf{LLM-SQL}, starting from an overly simple baseline (no column reordering) causes evolution to stagnate and yield poor hit rates, as the system fails to discover row-specific reordering. In contrast, starting from the published SOTA, which already achieves high hit rates, enables OpenEvolve to target the runtime bottleneck, evolving a solution in 100 iterations that is $3\times$ faster.

On the other hand, overly strong baselines that encode near-SOTA logic or rely on high-level APIs can limit the search to shallow micro-optimizations. In \textbf{Can't-Be-Late} (Section~\ref{sec:cant-be-late}), evolution from a simple greedy baseline produced better results than starting from the stronger Uniform Progress policy, which restricted exploration.

To navigate this trade-off, we recommend seeding evolution with three to five diverse baselines, potentially generated using coding assistants. We validate this with an ablation study on \textbf{LLM-SQL} using OpenEvolve. Seeding all islands with the same base program (the existing SOTA) caps combined scores at \(0.74\). In contrast, initializing islands with \emph{structurally diverse} programs, ranging from basic designs that capture the core column-reordering strategy to stronger heuristics that incorporate lightweight statistical signals (e.g., column-frequency and prefix-match patterns), allows the islands to start from different regions of the search space. This diversity enabled evolutions from one island to reach a combined score of \(0.7755\). Notably, only the runs that included these diverse seeds had a final score higher than 0.74. 

\takeaway{2}{Start evolution with diverse, domain-specialized seeds.}

\textbf{[Prompt Generator] More or fewer hints?} 
While detailed problem specifications are generally beneficial, the utility of solution hints (i.e., specific suggestions for how to approach the problem) is nuanced.
Too much guidance can risk \textit{premature convergence} and prevent the discovery of novel solutions, while too little can make the search inefficient.
For example, in \textbf{EPLB}, hints could have helped avoid wasted iterations on ``extreme'' replication strategies. Conversely, in \textbf{Transaction Scheduling}, hints about batching biased the search toward sub-optimal designs, whereas leaving it unconstrained enabled OpenEvolve to discover a 30\% faster greedy policy. 
This illustrates a fundamental trade-off: providing more hints accelerates by pruning the search space, whereas withholding hints fosters broader exploration, potentially at the cost of efficiency.

We recommend a two-phase approach to balance exploration and exploitation. In the first phase, we start with a generic, problem-only prompt with no specific algorithmic hints to allow the framework to explore freely. In the second, we inject targeted algorithmic hints into the prompt so that the search can escape local minima. We switch to this phase if the best combined score does not improve over a number of iterations.
In practice, this prompting approach effectively unblocks runs that would otherwise plateau. 
For both \textbf{EPLB} and \textbf{Cloudcast}, initial progress using generic prompts (``use optimization techniques'') plateaued after 12 and 8 iterations, respectively.
For EPLB, injecting a hint on proportional allocation strategies that guided the model to discover Hamilton apportionment, breaking the plateau (+28\%) and yielding +60\% total improvement. For CloudCast, suggesting a shared-tree multicast routing topology (replacing naive per-destination paths) led to a program that improved the score by +48\%. In contrast, the evolution that continued with the generic prompt failed to make further progress.

\takeaway{3}{Iteratively refine hints to balance the exploration-exploitation trade-off.}

\subsection{Evaluation: The Solution is Only as Good as the Evaluator}

\textbf{[Evaluator] Avoid overfitting.}
To prevent overfitting (i.e.,  generated solutions hard-code behaviors to narrow workloads), evaluation feedback should be robust. 
Specifically, we highlight two strategies. 
First, ensure \emph{test set diversity} via broad test sets that cover edge cases. Our ablation study on \textbf{Can't-Be-Late} (Appendix~\ref{appendix-cbl}) shows that limiting spot scheduling evaluation to a single availability pattern degraded performance on unseen workloads, whereas including diverse traces mitigated this issue. 
Second, employ \emph{validation safeguards}, such as separate validation sets and a Lines of Code (LOC) penalty to discourage hard-coded solutions.

\takeaway{4}{Enforce diverse test sets and safeguards.}

\textbf{[Evaluator] Prevent reward hacking.}
Robust evaluators are essential to prevent reward hacking in which generated solutions exploit loopholes to maximize scores without solving the problem.
For example, in \textbf{MAS}, some candidates bypassed one stage of the evaluation pipeline to achieve high scores without performing the full task. On \textbf{EPLB}, the LLM achieved high load balance by assigning zero weight to 97\% of experts. We later corrected this by explicitly enforcing that every expert must receive at least one replica.
To mitigate such failures, evaluators should combine multiple signals (e.g., correctness, efficiency, robustness) and include adversarial tests. This prevents the model from ``gaming'' the evaluation and aligns candidate solutions with intended constraints.

In particular, we observe that allowing full-file rewrites significantly increases the frequency and subtlety of reward hacking compared to scoped (i.e., diff-based) edits. In \textbf{Prism}, full rewrites on OpenEvolve enabled the LLM to delete the GPU memory limit variable, which would be immutable under diff-based edits (since it can be outside of the evolve block). The generated program caught the exception triggered by the undefined memory limit and continued execution, effectively bypassing the constraint. Similarly, in \textbf{Cloudcast}, where the model should only optimize find\_path(), full rewrites allowed it to omit broadcasting some data partitions, illegitimately reducing costs by 90\%.

Consequently, diff-based edits provide tighter control against reward hacking.
By restricting the LLM's ability to change code, we enforce ``correctness by construction'' and reduce the burden on the evaluator to catch correctness violations. Furthermore, limiting the number of changed LoC constrains deviation from the parent solution, analogous to KL-divergence penalties in RL (e.g., TRPO~\cite{schulman2015trust}, PPO~\cite{schulman2017proximal}) that stabilize policy updates.
Extending this logic, adjusting the edit scope allows us to regulate diversity, ensuring that child programs diverge sufficiently from their parents to drive exploration.

We recommend starting with diff-based edits and selectively transitioning to full rewrites when greater structural diversity is needed. To mitigate the increased risk of reward hacking during full rewrites, we advise refactoring the target program to remove code not needed for evolution. 

\takeaway{5}{Prevent reward hacking by restricting edits.}

\textbf{[Evaluator] Score carefully}. 
Having a precise scoring function is key to obtaining good evolution results. 
We recommend designing smooth reward functions to facilitate incremental search.
For instance, assigning penalized scores to invalid programs, rather than zeroing them out, preserves useful logic for future iterations (see "Learn from Failures").
Moreover, evaluation consistency across runs is crucial. For example, \textbf{EPLB} algorithm runtime is hardware-dependent, so we had to carefully isolate experiments to prevent interference from affecting measurements.

\takeaway{6}{Use smooth and deterministic scoring functions.}


\begin{table*}[t]
\centering
\small
\setlength{\tabcolsep}{4pt}
\renewcommand{\arraystretch}{1.2}
\begin{tabularx}{\textwidth}{@{}lrrrrrrrrrrrr@{}}
\toprule
\textbf{Framework} &
\textbf{EPLB} & \textbf{PRISM} & \textbf{TELEMETRY} & \textbf{TXN} &
\textbf{CLOUDCAST} & \textbf{NS3} & \textbf{CBL} & \textbf{CBL-Multi} &
\textbf{LLM-SQL} & \textbf{MAS} & \textbf{Overall Avg} \\
\midrule
\textbf{OpenEvolve} & 248 &  73 & 230 & 242 & 175 & 396 & 226 & 160 & 301 & 565 & \textbf{262} \\
\textbf{GEPA}       & 499 & 314 & 634 & 334 & 537 & $\sim$380 & 219 & 181 & 364 & 600 & \textbf{406} \\
\textbf{Shinka}     & 489 & 385 & 463 & 574 & 402 & 427 & 368 & 344 & 505 & 624 & \textbf{458} \\
\bottomrule
\end{tabularx}
\caption{LOC aggregated by framework (all models).}
\label{tab:loc-agg-framework}
\end{table*}

\begin{table*}[t]
\centering
\small
\setlength{\tabcolsep}{4pt}
\renewcommand{\arraystretch}{1.2}
\begin{tabularx}{\textwidth}{@{}lrrrrrrrrrrrr@{}}
\toprule
\textbf{Model} &
\textbf{EPLB} & \textbf{PRISM} & \textbf{TELEMETRY} & \textbf{TXN} &
\textbf{CLOUDCAST} & \textbf{NS3} & \textbf{CBL} & \textbf{CBL-Multi} &
\textbf{LLM-SQL} & \textbf{MAS} & \textbf{Overall Avg} \\
\midrule
Gemini-3 & 337 & 186 & 310 & 243 & 223 & 387 & 163 & 143 & 331 & 566 & \textbf{289} \\
GPT-5    & 487 & 328 & 574 & 523 & 519 & 415 & 378 & 314 & 449 & 626 & \textbf{461} \\
\bottomrule
\end{tabularx}
\caption{LOC aggregated by model (all frameworks).}
\label{tab:loc-agg-model}
\end{table*}

\textbf{[Generator] Select the right model.}
Given that models exhibit distinct ``thinking'' and coding styles that impact their effectiveness across different problems, we recommend conducting preliminary evolution runs with multiple models to select an effective LLM.
GPT-5 tends to produce longer, modular code with extensive edge case handling, whereas Gemini-3.0 generates shorter, minimal code with less abstraction. 
Tables~\ref{tab:loc-agg-framework} and \ref{tab:loc-agg-model} provide a detailed breakdown of these trends. Among frameworks, OpenEvolve generates the most concise implementations (averaging 262 LOC), roughly 1.5-1.7$\times$ shorter than GEPA (406 LOC) and Shinka (458 LOC). This framework-level efficiency compounds with model selection: Gemini-3.0 yields consistently shorter programs (289 LOC avg.) compared to GPT-5 (461 LOC avg.), which were $1.6x$ longer.
We analyze two key implications: modularity and program length. 

Modular programs are more effective for problems requiring the synthesis of multiple techniques.
For GEPA, GPT-5 outperforms Gemini-3.0 on all tasks requiring the combination of 
three or more techniques (\textbf{EPLB, Cloudcast, Transaction Scheduling}, Table~\ref{tab:cross-domain-techniques}), producing 3--4$\times$ more code and 3--8 $\times$ more helper functions.
Conversely, Gemini-3 excels on simpler algorithmic tasks (\textbf{CBL, Telemetry Repair}).

While program length does not inherently correlate with effectiveness, we recommend filtering overly long programs (e.g., by tuning the maximum character limit parameter in OpenEvolve). For instance, for \textbf{NS3} with ShinkaEvolve, GPT-5's complex code introduced memory safety bugs (e.g., use-after-erase, null dereference) and type confusion absent in Gemini's simpler and shorter implementations. 
Recent studies confirm that excessive length often signals unreliability~\cite{meyerson2025solvingmillionstepllmtask}, correlating with higher error rates and low-quality patterns~\cite{yu2025dapoopensourcellmreinforcement}. 

Across all problems (Tables~\ref{tab:loc-agg-framework} and \ref{tab:loc-agg-model}), OpenEvolve produces the shortest programs on average (262 LOC), followed by GEPA (406 LOC) and Shinka (458 LOC). 
Across models, Gemini-3 yields shorter programs on average (289 LOC) than GPT-5 (461 LOC).

\takeaway{7}{Select LLM based on solution composition.}

\subsection{Feedback: The ``Devil'' is in the Details}

\textbf{[Evaluator] Calibrate feedback granularity.}   
To efficiently guide the search, we recommend matching the feedback granularity to the problem structure. For problems for which the evaluation generates multiple results, we should provide per-result feedback (e.g., per-expert in \textbf{EPLB}, per-trace in \textbf{CBL}). 
This granular approach exposes workload-specific failure modes and guides improvements on under-performing cases. For example, GEPA’s ability to leverage per-result feedback enabled it to return the best programs in multi-result problems (\textbf{EPLB, CBL, CBL-Multi, Cloudcast}) compared to the two other ADRS frameworks that we evaluate.

However, as mentioned before, excessive feedback can be counterproductive. We perform an ablation study on \textbf{CBL}, measuring three levels of feedback: \emph{minimal} (aggregate cost only), \emph{moderate} (worst five scenarios), and \emph{detailed} (per-trace multi-dimensional breakdowns). Moderate feedback achieves the best cost reduction (13.0\%), outperforming minimal (7.7\%) and detailed (10.2\%) feedback; this confirms that while actionable guidance helps, too much detail can lead to overfitting.

\takeaway{8}{Calibrate feedback granularity to provide actionable guidance without causing overfitting.}

\textbf{[Evaluator] Learn from failures.}
Rather than discarding failed programs, it can be useful to retain them (i.e., via appropriately low scores or repair loops) as they might contain good ideas that contribute to future improvements.
We validate this through an ablation study that implements an automatic retry mechanism in OpenEvolve. In \textbf{Transaction Scheduling}, repairing a child program that failed with a Python syntax error improved the final score by 1.1$\times$. 
Similarly, for \textbf{LLM-SQL}, a candidate causing a 600s timeout was repaired into a 2.6s variant that became the run's best program.
These cases illustrate that first-attempt failures can yield high-quality descendants if given a chance for consideration and refinement. 
We note that the general trade-off, rich feedback for faster convergence, and prematurely narrowing the search space apply here as well.

\takeaway{9}{Learn from failures to gain insights for evolution.}

\textbf{[Evaluator] Leverage meta-feedback selectively.} 
Incorporating meta-feedback from evolutionary history can accelerate convergence in some cases. 
Frameworks like ShinkaEvolve extract insights during evolution by periodically summarizing recent successful strategies, generating recommendations with additional model calls, and using these to tune subsequent prompts. 
On \textbf{Prism}, the meta-feedback helped identify when greedy and local search strategies stagnated, explicitly steering the model toward a more effective binary search solution. 
Conversely, on \textbf{CloudCast}, the meta-feedback provided incorrect guidance: it recommended ``edge reuse to minimize cost,'' overlooking the fact that concurrent use of network edges creates congestion, which increases transfer time and leads to lower scores.

\takeaway{10}{Meta-feedback helps guide search when accurate.}

\textbf{[Evaluator] Handle non-deterministic evaluation with care.}
Non-deterministic evaluators, where identical code yields varying scores across runs, amplify framework-specific trade-offs. For instance, GEPA's subsample filtering with noisy partial comparisons can lead to rejection of valid proposals due to variance in scores. Shinka's meta-learning can lock in early noise, mistaking randomness for signal that can bias future generations. OpenEvolve was most resilient on our non-deterministic use cases (\textbf{NS3, Telemetry Repair}), utilizing population diversity to average out fitness noise and diff-based constraints to prevent data overfitting.
On \textbf{NS3}, where scores can vary due to randomized flow start times within a 10ms, OpenEvolve's population approach smoothed the signal over runs while Shinka's early lock-in led to high variance between candidate solutions and GEPA rejected 89–97\% of proposals. On \textbf{Telemetry Repair}, where input measurements contain synthetic errors, OpenEvolve's diff-based edits led to simpler code that generalized better over noisy data.

(simulator noise), Shinka's early lock-in created a \textbf{14.71 point variance} between identical runs, while GEPA rejected \textbf{89–97\% of proposals}; OpenEvolve's population approach successfully smoothed the signal. On \textbf{Telemetry Repair} (data noise), OpenEvolve's simplicity pressure produced significantly shorter code (\textbf{174 LOC} vs GEPA's 868 LOC) that generalized better. We recommend \textbf{delaying meta-feedback} and \textbf{using median scores} from multiple evaluations to mitigate these risks. Beyond framework selection, we recommend two additional mitigations: (i) run more iterations before applying meta-feedback, and (ii) use median scores from multiple evaluations of each candidate.

\takeaway{11}{Encode resilience for non-deterministic evaluators.}

\section{Limitations and Open Challenges}

In this section, we characterize the limitations of \SYS/, distinguishing between the types of problems for which the approach is well suited and those for which it is not. We then outline several open challenges aimed at improving \SYS/ so that it can address a broader range of problems.

\subsection{Which Problems Are Best Suited?}
\label{sec:limitations}
Our experience suggests that existing \SYS/ frameworks excel on system problems with three key properties:
\begin{itemize}

\item \emph{Isolated changes:} 
\SYS/ works best when improvements can be made in a small, self-contained part of the system (e.g., schedulers, cache managers, load balancers).
In contrast, problems that require coordinated changes across multiple modules, e.g., distributed protocols with interacting state machines~\cite{paxos,raft}, remain difficult due to limited multi-file reasoning and limited context lengths.

\item \emph{Reliable evaluations:} 
\SYS/ requires evaluators that can unambiguously rank candidate solutions based on correctness and performance. This is straightforward when correctness relies on clear metrics or invariants (e.g., load balance factor).
It is far harder when semantic equivalence is difficult to verify, such as in arbitrary database query plan rewrites, where checking equivalence is generally undecidable~\cite{AbiteboulHullVianu1995}.

\item \emph{Efficient evaluations:} 
Since evolution may require thousands of iterations, each evaluation must be cheap. Problems that require hours of GPU time or large distributed testbeds (e.g., weight compression, large-scale training; see Section~\ref{sec:case_studies}) make evolution prohibitively slow or expensive.
\end{itemize}

As such, we do not expect \SYS/ to be as effective for problems that can be cast as optimization problems and solved directly using existing solvers, such as integer linear programming (ILP) solvers. 

\subsection{Open Challenges}
\label{sec:open_challenges}
As \SYS/ emerges as a promising approach to accelerate systems research, two natural questions arise: what should we build to better support \SYS/, and how should we improve \SYS/ itself?

\subsubsection{Supporting \SYS/ with Better Evaluators}
\label{sec:better-evaluators}

\SYS/ frameworks are only as good as the evaluators guiding them. Successful discovery requires evaluators that provide accurate, fast, and detailed feedback:

\begin{itemize}

\item \emph{Fidelity:} Evaluators must capture the salient system behaviors that are relevant to the problem being solved (e.g., flow-level or packet-level fidelity in networking). Crucially, the desired level of fidelity depends not only on the system under study but also on the solution space that \SYS/ explores.

\item \emph{Generality:} Evaluators must support diverse workloads to provide robust feedback signals and prevent overfitting to a single configuration or dataset.

\item \emph{Speed and reliability:} Rapid evaluation is key to \SYS/ scalability. Building infrastructure that supports rapid forking and rollback (e.g., lightweight VMs, container snapshots, or database cloning) can drastically accelerate feedback cycles~\cite{liu2025supporting}.

\end{itemize}

Next, we highlight two directions for improving efficiency.

\textbf{Problem-specific simulators.} Simulators offer fast, low-cost evaluation but are hard to design, as they must balance fidelity and simplicity. 
A practical solution is to build domain-specific simulators that capture only behaviors relevant to the task (e.g., modeling CPU scheduling without full operating system details). 
Tools like OpenEvolve can iteratively refine these simulators until their target metrics (e.g., latency, throughput) align with real systems. Though costly to build, such simulators can be reused across related problems, amortizing development effort.

\textbf{Cascading evaluators.} 
Fast and accurate evaluation can be achieved via cascading evaluators: progressing from fast, coarse-grained cost models to slower but high-fidelity simulators.
For example, one might begin with a simple cost model (e.g., for database queries~\cite{SiddiquiJindalQiaoPatelLe2020}), then progress to simulators of increasing granularity, followed by emulators, and finally tests on the real system. In networking, for instance, session-level simulators offer coarser evaluation compared to packet-level simulators. This mirrors standard research practice: prototype quickly, then validate precisely. 

\subsubsection{Improving \SYS/}
To extend the reach of \SYS/, we need advances across the following key components.  

\textbf{Problem specification and prompting.}
Existing \SYS/ frameworks often rely on simple prompts that provide only a problem description.
Just as human researchers ground their ideas in prior work, \SYS/ frameworks should incorporate retrieval techniques~\cite{packer2023memgpt} to draw from a broader body of knowledge (e.g., academic literature, documentation, and related examples) to better guide their search.
Prompt evolution~\cite{alphaevolve, agrawal2025gepa}, which allows the LLM to refine its own context, can help manage the trade-off between exploration and exploitation in providing hints by starting with generic guidance and then dynamically incorporating more targeted hints or examples when search stagnates.

\textbf{Solution generator.}
\label{sec:future-work-sol-generator}
A strong solution generator should act like an autonomous developer, capable of navigating, reasoning over, and modifying the full system codebase rather than isolated code snippets. 
For example, modern AI workloads often rely on workload analysis and cross-layer optimizations to reduce hardware costs~\cite{chung2024toward}. Similarly, optimizations for distributed communication protocols require coordinated changes to both sender and receiver logic~\cite{raft,opaxos}. 
Achieving this demands further agentic capabilities: understanding dependencies, invoking analysis tools, and reasoning coherently across non-contiguous code modules.
Beyond a single model, \SYS/ frameworks support ensembles of specialized agents that can be dynamically composed based on the given problem, much like forming a research team with complementary skills.

\subsubsection{Solution Selector and Search Process}
The discovery process of ADRS frameworks remains largely a black box. Current evolutionary searches are monolithic, often producing a single final program that can mix good ideas with poor implementation and thus penalize promising concepts. Prior work~\cite{FunSearch2024} suggests separating ideation from code generation to address this issue. 

Moreover, evolutionary search is often inefficient, frequently looping over failed heuristics or repeated errors (Appendix~\ref{app:failures}). We need more flexible frameworks that support finer-grained feedback, enabling humans or LLMs to lock in working code, boost diversity to escape local optima, or roll back to prior versions when evolution stalls.

\subsubsection{Objectives, Feedback, and Preference Learning}
Some ADRS frameworks, such as OpenEvolve, require researchers to formalize intuitive trade-offs into numerical weights, which can be difficult in practice. For example, in \textbf{EPLB}, we struggled with how to weigh the importance of load balance against algorithm runtime. 
Furthermore, feedback granularity, which is currently determined by users, also shapes evolution: our \textbf{Can't-Be-Late} ablation confirms that moderate, scenario-level feedback outperforms both minimal (weak guidance) and detailed (overfitting-prone) feedback. 

Future systems could instead learn user preferences automatically. One approach is \emph{preference learning}: ADRS could present researchers with various solutions (e.g., “Solution A is faster but less fair; Solution B is fairer but slower”) and infer the underlying objective function from user choices. Another approach is \emph{inverse reinforcement learning (IRL)}: ADRS could infer reward functions from user-provided examples of ``good'' solutions.


\textbf{Framework-level changes.}
\label{sec:future-work-framework}
Finally, we discuss open challenges related to the overall \SYS/ framework. 

\textit{Hyperparameter automation.} Balancing exploration and exploitation remains difficult and currently relies on trial and error. Future work should focus on automating this tuning process. For instance, a meta-learning layer could allow the system dynamically adjust during evolution, making the entire framework more reliable and accessible.

\textit{Human-\SYS/ interaction.} 
The optimal balance between synchronous (interactive assistants like Cursor) and asynchronous (autonomous frameworks like OpenEvolve) user interfaces remains an open question. A key challenge is defining when human guidance adds values versus when \SYS/ should act autonomously. 


\section{\SYS/'s Impact on the Research Process}
\label{sec:esearch-process-impact}

Despite their limitations (see Section~\ref{sec:limitations}), \SYS/ frameworks can help researchers in two key ways:

\begin{itemize}
\item \emph{Accelerate discovery:}
\SYS/ frameworks automate tedious tasks, such as implementation and, to some extent, debugging, freeing researchers to focus on problem selection and system design. Even imperfect solutions are useful, as they often reveal new directions; in our telemetry repair case study, AI-generated insights helped guide a better human-designed algorithm.

\item \emph{Achieve better-than-human results:}
\SYS/ tools can explore the solution space more thoroughly. While humans often stop after a breakthrough, AI can continue testing variations with compounding incremental gains. In \textbf{EPLB}, OpenEvolve produced an algorithm that surpassed SOTA by exploiting potentially overlooked optimizations.
\end{itemize}

As such, \SYS/ has the potential to reshape systems research. One way to think about \SYS/ is as providing researchers with a virtually unbounded number of ``assistants.'' Much like junior researchers, these frameworks are most effective when given clear specifications and well-defined goals.

Consequently, \SYS/ adds another layer to the existing research hierarchy. In academia, for example, faculty members typically advise Ph.D.\@ students or postdocs, who in turn mentor undergraduate students. Each of these roles can now leverage \SYS/ tools to accelerate their work and broaden the scope of what they can accomplish.

This shift raise a natural question: \textbf{how will the role of a researcher change?}
As \SYS/ tools become more effective, we believe researchers will have more time to focus on higher-leverage activities, namely, choosing problems and formulating them precisely. If successful, \SYS/ has the potential to elevate the entire research hierarchy, faculty and students alike, and make them far more productive.

\section{Related Work}
\label{sec:related-work}


\SYS/ builds on a long line of research that combines large-scale search with AI to tackle complex problems. 

\textbf{Pre-LLM AI for System Optimizations.}
Prior to LLMs, machine learning had already been widely applied to optimize systems.
In databases, learned models addressed query optimization~\cite{Marcus_2019}, cardinality estimation~\cite{balsa, NeuroCard}, learned indexes~\cite{kraska2018caselearnedindexstructures}, and automated system tuning~\cite{vanaken2017automatic}.
Reinforcement learning (RL) and other techniques advanced core networking problems, including congestion control~\cite{jay2018internet}, packet classification~\cite{liang2019neuralpacketclassification}, and topology modeling~\cite{Rusek_2020}. 
More broadly, RL has been applied in systems for scheduling over data processing workloads~\cite{Mao2019Decima}, physical device placement~\cite{Mirhoseini2017DeviceICML}, and video streaming~\cite{Du2020ServerDriven}.

\textbf{Automated discovery with learned approaches.} AI has increasingly powered automated discovery in complex domains. Early successes include AlphaGo~\cite{alphago} and AlphaZero~\cite{silver2018alphazero}, which demonstrated how search and RL can master games, and AlphaFold~\cite{alphafold}, which achieved breakthroughs in protein structure prediction. 
AlphaDev~\cite{alphadev} extended this to low-level algorithms, while AlphaChip~\cite{alphachip2024} uses RL for chip layouts, and Big Sleep~\cite{bigsleep} applies AI agents to detect software vulnerabilities. More recently, benchmarks, such as AlgoTune~\cite{press2025algotune}, have been developed to evaluate LLM program optimization abilities.

\textbf{LLM-based coding assistants.} Tools like GitHub Copilot~\cite{GitHubCopilot}, Cursor~\cite{CursorAgent2024}, Codex~\cite{OpenAICodex}, and Claude Code~\cite{ClaudeCode2025} accelerate research by helping researchers rapidly prototype ideas, build simulators, and implement baselines. 
They enable rapid translation of high-level concepts into working implementations. 
Recent work also explores using LLMs for performance-critical code~\cite{hong2025autocompllmdrivencodeoptimization}, such as GPU kernel code generation and optimization~\cite{ouyang2025kernelbench}, further illustrating their potential for \SYS/. 

\textbf{LLM-driven research.}
Beyond coding, recent work leverages LLMs to automate broader parts of the research process. 
Frameworks such as AlphaEvolve~\cite{alphaevolve, nadga2025alphaevolve} and OpenEvolve~\cite{openevolve} use the MAP-Elites algorithm and island models to evolve new algorithms; GEPA~\cite{agrawal2025gepa} employs reflective prompt evolution for better LLM generation; and LLM4AD~\cite{liu2024llm4ad} provides a unified platform for algorithm design. 
Others include ShinkaEvolve~\cite{shinkaevolve} (sample-efficient evolution), EvoPrompt~\cite{evoprompt} (genetic prompt optimization), MetaMuse~\cite{ma2025algorithmgenerationcreativeideation} (self-reflective generation), and PolicySmith~\cite{dwivedula2025manmade} (solution selection). Glia~\cite{hamadanian2025gliahumaninspiredaiautomated} presents an agentic workflow for systems design, reporting strong results on distributed LLM inference.

Attempts at end-to-end research automation are also emerging. MLGym~\cite{nathani2025mlgym} offers AI research agent benchmarks, while Code Researcher~\cite{singh2025code} explores agents for large-scale software engineering. Work on self-evolving AI agents~\cite{fang2025comprehensive} and broader systems applications~\cite{liang2025nexthorizon,zhouneuripstalk} is also growing. 
Darwin~\cite{zhang2025darwin} demonstrates open-ended, self-referential code improvement, automatically evolving stronger coding agents.

Our work focuses on automated algorithm discovery in systems, where strong evaluators enable the reliable verification needed for productive automation.

\section{Conclusion}
\label{sec:conclusion}
As ADRS automates algorithm discovery, the role of the human researcher will evolve. Much like how an academic advisor guides a student, the researcher of the future will direct these AI frameworks by defining problems, steering the research process, and critically evaluating the results. 
One of the most profound implications of this shift is the potential for a \emph{virtuous cycle}. We can use an \SYS/ to improve itself. As recent work has shown~\cite{zhang2025darwingodelmachineopenended}, models can learn to refine their own reasoning, debug code, and discover more effective strategies. 
By rapidly iterating on their own capabilities, AI agents will compound the pace of scientific discovery.

In this paper, we have demonstrated the potential of \SYS/ in systems research. Our case studies show that the \SYS/ approach can already outperform human baselines on key performance problems. While it is still very early, we call on the systems community to embrace these tools, not just as accelerators of research, but as subjects of research. Improving \SYS/---making it efficient, scalable, and reliable---is itself a systems challenge. As such, we believe that system builders are uniquely positioned to shape the future of AI-driven discovery.



\bibliographystyle{plainnat}
\bibliography{references}

\newpage
\newpage
\appendix

\clearpage
\section{A Pilot Survey of Time Spent in Systems Research}
\label{app:survey_result}
\accheng{@Ion, do we want to keep this still?}

\begin{figure}[H]
  \centering
    \includegraphics[width=0.45\textwidth]{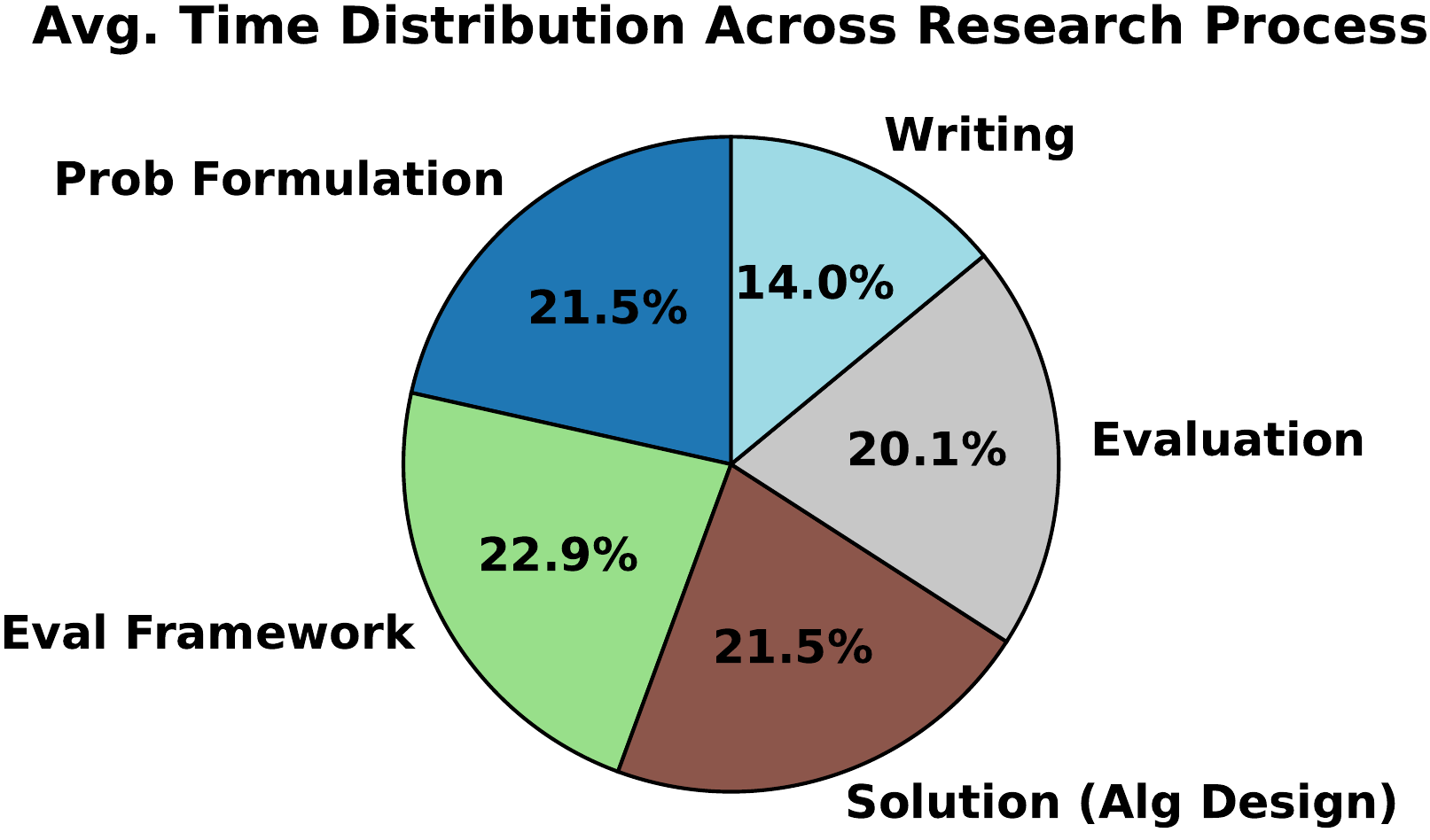}
  \caption{Time spent in various stages of the systems research process in systems, based on a survey of 31 PhD students. \textit{Algorithm Design} (21.5\%) and \textit{Evaluation} (20.1\%) together account for over 40\% of total effort, highlighting a significant opportunity for leveraging AI to accelerate this process.}
  \label{fig:research-process-survey}
\end{figure}

To understand how researchers allocate their time across stages of the systems research process, we conducted a small survey of 31 PhD students in broadly systems research at a US university. Figure~\ref{fig:research-process-survey} shows the approximate fraction of effort they report spending in each stage.



\section{Additional Case Studies}

We also provide descriptions on the other case studies we run, including network telemetry repair, cloudcast, global model placement, multi-agent system optimization, and datacenter congestion control.

\subsection{Case Study \#5: Telemetry Repair}
This task identified by HotNets’24~\cite{krentsel2024case} studies repair of faulty router telemetry signals, which can become buggy due to router faults or collection infrastructure errors. Such inconsistencies (for example, counters on the two ends of a link not matching) can cause the network controllers to make incorrect decisions. The objective is to detect and repair faulty telemetry to produce a self-consistent view of network state.  

We use multiple evolve frameworks (e.g. OpenEvolve, GEPA, ShinkaEvolve) to evolve repair strategies, running 100 iterations  with GPT-5 and Gemini 3.0-Pro models, supplemented by contextual hints from the HotNets’24 paper. The best evolved program introduces structured repair logic, including averaging nearby counters to reduce noise, and separating repair and confidence estimation into distinct steps. It achieves a repair score of 95\% and confidence calibration of 95\%, outperforming the HotNets’24 solution (86\% repair, 65\% confidence).

\subsection{Case Study \#6: Cloudcast}
The Cloudcast problem, published in NSDI ’24~\cite{wooders2024cloudcast}, studies cost-aware multicast across multi-region and multi-cloud topologies. The objective is to construct an overlay connecting a source, multiple destinations, and optional waypoints so as to minimize egress cost.  

Our initial program is a direct replication strategy: the source sends data independently to each destination. While simple, this approach often incurs high egress costs when destinations are located in distant or expensive regions. To guide evolution, the evaluator tests candidate algorithms across 10 multi-region, multi-cloud configurations and assigns a total score based on the overall egress cost of the scheduled paths. We use our evolved frameworks to run 100 iterations in about one hour, costing less than \$10 for the entire run.  

The evolved solution successfully discovers a Steiner tree strategy, achieving an average cost reduction of 31.1\% compared to naive direct replication. The best performing solution is close to the human state-of-the-art solution. This solution constructs a cost-efficient multicast tree by introducing intermediate waypoints. For example, data may be replicated once at a cheaper waypoint region and then forwarded to multiple destinations, reducing the total egress cost.

\subsection{Case Study \#7: Model Placement (Prism)} The research problem~\cite{yu2025prism} focus on the challenge of multi-LLM serving on shared GPUs under bursty, heterogeneous workloads. The key metric is the KV pressure ratio (KVPR), defined as the SLO-weighted request rate divided by available KV cache memory. The optimization goal is to minimize the maximum KVPR across GPUs, thereby reducing contention and improving SLO compliance.  

The base program is the algorithm from the paper~\cite{yu2025prism}: models are placed sequentially onto the first GPU with sufficient remaining memory, without considering long-term balance. While feasible, this approach often overloads some GPUs while leaving others underutilized.  
We use OpenEvolve to evolve improved placement strategies, guided by a scoring function that combines execution correctness with the KVPR objective. The evolution runs with GPT-5 and Gemini-3. The best evolved program achieves an 18.5\% higher score compared to the state-of-the-art reported in the original paper.  

The evolved strategy mirrors the SOTA algorithm from the paper and assignment used in PRISM, but crucially adds a local improvement stage. After the initial placement, it repeatedly tests whether moving or swapping models between GPUs reduces the maximum KVPR, applies such refinements, and stops once it finds a move that can reduce the current maximum KVPR. 



\subsection{Case Study \#8: Multi-Agent System Optimization (MAS)} 
The research problem extends the work of~\cite{cemri2025multi} (NeurIPS'25), which studies diagnosing and repairing failures in multi-agent LLM systems (MAS) and proposes MAST as the taxonomy of MAS failure modes. Such systems (e.g., MetaGPT~\cite{metagpt}, ChatDev~\cite{chatdev}) often suffer from breakdowns in coordination, memory, or communication that reduce task success. The challenge is how to improve multi-agent systems, evolving more robust agent architectures, prompts, and inter-agent communication patterns to increase reliability and performance.  

The base program is a direct adaptation of MetaGPT~\cite{metagpt}, assembled into a minimal Python implementation (roughly 400 LOC). This initial version defines fixed agent roles, communication protocols, and system prompts, but frequently encounters the failure modes identified in the MAST. To guide evolution, the evaluator uses the MAST annotator, which assigns a score of $1/(1+\text{total FM occurrence})$, penalizing common coordination and memory failures.  

We ran three evolution configurations with different mutation scopes: (1) modifying agent definitions and communication schemes, (2) introducing new verification and communication flows using GPT-5, and (3) allowing changes to the number and types of agents, i.e., the system topology. The evolved solutions introduced innovations such as improved context management (v1) and verification/communication flows (v2), though removing verification in v3 degraded performance. Overall, downstream program development success improved from 40\% in the base program to 47\% (v1) and 53\% (v2) on the ProgramDev-v1 benchmark, before dropping to 30\% in v3. The fast that verification agent was removed in v3 was an example of reward hacking (since we penalize the verification failures, the evolution algorithm got rid of the whole verification when it could) and an example of the importance of carefully tuning which parts of the initial code the evolution algorithm is allowed to change. These results show that the evolve frameworks can automatically discover MAS design refinements that improve robustness, though careful control over mutation scope is critical to avoid reward hacking.

\subsection{Case Study \#9: Datacenter TCP Congestion Control (NS3)}
This research problem extends PowerTCP~\cite{powertcp} (NSDI `22), which proposes a delay-based congestion control protocol for datacenter networks that uses a ``power'' signal (i.e., the ratio of throughput to delay) to detect congestion. Traditional datacenter TCP variants struggle with incast scenarios where many flows simultaneously converge on a single receiver, causing buffer bloat and packet loss. We leverage the existing benchmark implementation in Network Simulator 3 (NS-3)~\cite{ns3} to investigate the ability of ADRS frameworks to optimize networking applications written in C++. The challenge is how to improve congestion control algorithms, evolving more responsive rate adaptation mechanisms that maximize throughput while minimizing switch queue occupancy.

The base program is $\theta$-PowerTCP, a variant that does not require in-network telemetry and instead infers congestion purely from end-to-end RTT measurements. This initial implementation (roughly 400 LOC) defines the rate control logic, which calculates a power signal from RTT gradients and adjusts sending rates accordingly. The benchmark runs on NS3, where scores can vary due to randomized flow start times within a 10ms window, simulating a 10:1 incast scenario where ten flows are launched toward a receiver already handling a long-running background flow. To guide evolution, the evaluator uses a fitness function $e^{(\text{throughput}{\text{Gbps}} - 20)} - \text{queue\_length}/100$, rewarding high throughput while penalizing queue buildup.

The evolved solutions discovered several key innovations: (1) conservative flow startup, where new flows begin at 1/10 of the maximum link rate rather than full speed, preventing burst-induced congestion; (2) explicit congestion notification handling with forced minimum congestion signals; and (3) adjusted rate control parameters. The best evolved program reduced average queue length by 49\% compared to the original implementation while maintaining nearly identical throughput. It is worth noting that the most impactful optimization was the conservative initial rate, a counterintuitive finding since aggressive startup seems optimal for throughput but actually triggers severe congestion before the control loop can react. These results demonstrate that the evolution frameworks can automatically discover protocol-level optimizations in low-level systems code that might otherwise be overlooked.

\section{Additional Analysis}
\label{appendix-add}

We provide additional analyses that complement the main results and offer a deeper view of how ADRS behaves across settings and frameworks. We examine how changing train-set coverage affects policy quality, compare program length across frameworks and models, and present a schema table that formalizes the components of an ADRS setup. We also summarize recurring failure patterns observed and highlight cross-domain techniques that emerged during evolution.

\subsection{Changing Train Set Coverage}
\label{appendix-cbl}
\begin{figure}[h!]
  \centering
  \begin{subfigure}[t]{0.48\textwidth}
    \centering
    \includegraphics[width=\textwidth]{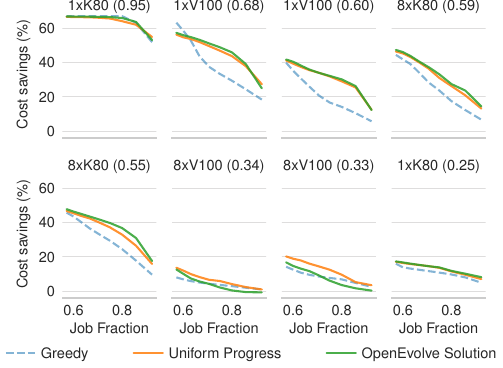}
    \caption{Training with 3\% of the available training set.}
    \label{fig:failure-taxonomy-a}
  \end{subfigure}
  \hfill
  \begin{subfigure}[t]{0.48\textwidth}
    \centering
    \includegraphics[width=\textwidth]{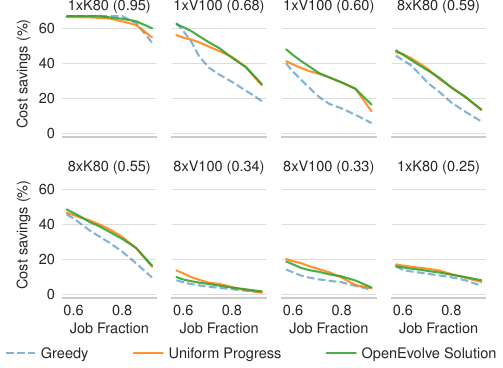}
    \caption{Training with the full training set.}
    \label{fig:failure-taxonomy-b}
  \end{subfigure}
  \caption{Impact of training-set coverage on policy evolution. 
  We split the data into 30\% training and 70\% testing. 
  The left panel uses 3\% of the training data, while the right panel uses the full set.}
  \label{fig:failure-taxonomy}
\end{figure}

We evaluate how training-set coverage affects the learned policies in the \textbf{CBL} case study. 
Figure~\ref{fig:failure-taxonomy} compares results obtained when training on 3\% of the training set versus the full training set.

Figure~\ref{fig:failure-taxonomy} shows results across all hardware profiles. 
Each panel reports cost-savings curves for Greedy, Uniform Progress, and the OpenEvolve solution as a function of job fraction. 
In both settings, the curves exhibit similar shapes, and the ordering among the three methods remains visually consistent across the two training-coverage conditions.

\subsection{Schema Table}
\begin{table*}[h]
\centering
\small
\begin{tabular}{p{0.25\textwidth}p{0.68\textwidth}}
\toprule
\textbf{Dimension} & \textbf{Key Components} \\
\midrule
\textbf{Prompt Generator (Problem Setup)} 
& \textit{Problem description}: problem domains -- computer systems (networking, distributed systems, databases, MLSys, etc.); problem description -- performance optimization, e.g., find the most cost-effective transfer graph \\
&\\
& \textit{Optimization objective}: e.g., latency, throughput, cost, algorithm runtime \\
& \textit{Constraints}: e.g., latency SLOs \\
\midrule
\textbf{Solution Generator} 
& \textit{LLM type}: what model used for solution generation, this includes reasoning vs.\ non-reasoning, tool-use vs.\ non-tool-use, LLM ensemble\\
&\\
& \textit{Number of iterations}: number of rounds to iterate the solution\\
\midrule
\textbf{Evaluator}
& \textit{Environment and test data}: testbed environment such as CPU simulator, database, GPU environment; test data and traces \\
&\\ 
& \textit{Initial Program}: the initial program to start with, use public source (GitHub/paper) if available, otherwise use simple algorithm \\
& \textit{Additional Baselines}: additional baselines for comparison if any \\
&\\
& \textit{Evaluator Feedback}: execution score, more advanced if any (e.g., error messages, human- or agent-in-the-loop feedback) \\
\midrule
\textbf{Solution Selector} 
& \textit{Selection algorithm}: e.g., greedy, random, or island algorithm \\
\specialrule{1pt}{0.4em}{0.4em} 
\textbf{How We Analyze Result} 
& \textit{Performance}: compare evolved result vs.\ initial program and SOTA algorithm \\
& \textit{Cost-benefit}: compute budget, LLM calls, simulation time (survey: was the quality gain worth the cost?)\\
& \textit{Other metrics}: robustness, generalization \\
\midrule
\textbf{How We Analyze Evolution Process} 
& \textit{Search trajectory}: from initial program to checkpoints to final outputs (examples of key transitions, what features are added) \\
&\\
& \textit{Common patterns}: where models get stuck, recurring failures \\
& \\
& \textit{Feedback granularity and utility}: scores, constraint violations, trace-level logs; which kinds of feedback resolve which failure patterns\\
\bottomrule
\end{tabular}
\caption{Expanded schema for problem formulation and evaluation of AI-driven algorithm discovery. Each row corresponds to an element in the setup. \shu{do we still need this table?} \accheng{cite in S5? need to double check}}
\label{tab:expanded-schema}
\end{table*}

Table~\ref{tab:expanded-schema} summarizes the components used to define and evaluate an ADRS setup. 
It covers four main elements---Prompt Generator, Solution Generator, Evaluator, and Solution Selector---and lists the specific factors included in each (e.g., problem description, optimization objectives, model choice, test environment, and feedback signals). 
The table also includes dimensions used in our analysis of results (e.g., performance and robustness) and of the evolution process (e.g., search trajectory, recurring failure patterns, and feedback utility). 
Together, these dimensions outline the information required to specify an ADRS problem and to evaluate both the resulting solutions and the evolution dynamics.

\subsection{Failure Patterns}
\label{app:failures}
\begin{table*}[!h]
\centering
\renewcommand{\arraystretch}{1.35} 
\setlength{\tabcolsep}{6pt} 
\resizebox{\linewidth}{!}{%
\begin{tabular}{p{2.8cm} l p{8.8cm}}
\toprule
\textbf{Category} & \textbf{Failure Type} & \textbf{Description} \\
\midrule
\multirow{2}{*}{\makecell[l]{Runtime\\Errors}} 
& \textit{Syntax \& Interface Errors} & Candidate solution fails to compile or integrate with evaluator. \\
& \textit{Budget Exhaustion} & Candidate exceeds resource limits (e.g., context window, API quotas, timeouts). \\
\addlinespace
\midrule
\multirow{3}{*}{\makecell[l]{Search\\Failures}} 
& \textit{Premature Convergence} & Search settles on a local optimal solution too early. \\
& \textit{Stuck-in-the-Loop} & Search repeats similar solutions without meaningful progress. \\
& \textit{Mutation Drift} & Search produces contradicting or random edits to the solution. \\
\addlinespace
\midrule
\multirow{4}{*}{\makecell[l]{Algorithm\\Failures}} 
& \textit{Misaligned Objectives} & Solutions ignore key constraints (e.g., latency SLOs). \\
& \textit{Sub-Optimal Optimizations} & Shallow changes (e.g., API calls) instead of substantive algorithmic improvement. \\
& \textit{Overfitting} & Hard-coded / narrow solutions underperform on unseen traces. \\
& \textit{Reward Hacking} & Solution exploits loopholes in the evaluator rather than solving intended problem. \\
\bottomrule
\end{tabular}
}
\caption{Common failure patterns in \SYS/ pipelines (distribution estimated from 420 LLM-judged traces).}
\label{tab:f}
\end{table*}

Table~\ref{tab:f} groups failures observed in \SYS/ pipelines into runtime errors, search failures, and algorithm failures. 
Runtime errors include syntax or interface issues and cases where candidates exceed resource limits. 
Search failures include premature convergence, repeated solution loops, and mutation drift. 
Algorithm failures include misaligned objectives, shallow optimizations, overfitting, and reward hacking. 
These categories reflect the types of failure modes observed across 420 LLM-judged traces.

\clearpage
\newpage

\onecolumn
\section{Config Files for ADRS Frameworks}
\label{sec:config_files}

We provide the configuration files we use for OpenEvolve, ShinkaEvolve, and GEPA. 

\subsection{OpenEvolve}
\begin{lstlisting}
# OpenEvolve Island-Based Evolution Configuration
# This configuration demonstrates the proper use of island-based evolution

# General settings
max_iterations: 100
checkpoint_interval: 10
log_level: "INFO"
random_seed: 42

# LLM configuration
llm:
  primary_model: "gpt-5"
  api_base: "https://api.openai.com/v1"
  api_key: ${OPENAI_API_KEY}
  primary_model_weight: 1.0
  temperature: 0.7
  top_p: 0.95
  max_tokens: 32000
  timeout: 600

# Database configuration with proper island settings
database:
  population_size: 100
  archive_size: 20

  # Island-based evolution settings
  num_islands: 5                   # Number of separate populations
  migration_interval: 5            # Migrate every 50 generations
  migration_rate: 0.1               # Migrate 10% of top programs

  # Selection parameters
  elite_selection_ratio: 0.1
  exploration_ratio: 0.3
  exploitation_ratio: 0.7
  # Note: diversity_metric fixed to "edit_distance"

  # Feature map dimensions for MAP-Elites
  # Default if not specified: ["complexity", "diversity"]
  # Comment out the line below to use the defaults
  # feature_dimensions: ["complexity", "diversity"]
  feature_bins: 10
  # Can also use per-dimension bins:
  # feature_bins:
  #   performance: 20
  #   correctness: 10

# Prompt configuration
prompt:
  num_top_programs: 3
  num_diverse_programs: 2
  use_template_stochasticity: true

# Evaluator configuration
evaluator:
  timeout: 300
  max_retries: 3
  cascade_evaluation: false
  parallel_evaluations: 4

# Evolution settings
diff_based_evolution: true
allow_full_rewrites: false
max_code_length: 60000
\end{lstlisting}

\subsection{ShinkaEvolve}
\begin{lstlisting}
evo_config:
  _target_: shinka.core.EvolutionConfig
  patch_types:
    - "diff"
    - "full"
    - "cross"
  patch_type_probs:
    - 0.6
    - 0.3
    - 0.1
  num_generations: 100
  max_parallel_jobs: 10
  max_patch_resamples: 3
  max_patch_attempts: 3
  llm_models:
    - "gemini-2.5-pro"
    - "gemini-2.5-flash"
    - "gpt-4.1-mini"
    - "gpt-4.1-nano"
    - "bedrock/us.anthropic.claude-sonnet-4-20250514-v1:0"
    - "o4-mini"
  llm_dynamic_selection: ucb
  llm_kwargs:
    temperatures:
      - 0.0
      - 0.5
      - 1.0
    max_tokens: 16384
  meta_rec_interval: 10
  meta_llm_models:
    - "gpt-4.1"
  meta_llm_kwargs:
    temperatures:
      - 0.0
  embedding_model: "text-embedding-3-small"
  results_dir: ${output_dir}
  
\end{lstlisting}

\subsection{GEPA}
\begin{lstlisting}
reflection_minibatch_size: 3 
max_iterations: 100 
reflection_lm: openai/gpt-5 # gemini/gemini-3-pro-preview
\end{lstlisting}

\end{document}